\documentclass[journal]{IEEEtran}
\usepackage{cite}
\usepackage{booktabs}
\usepackage{amsfonts} 
\usepackage{amssymb}
\usepackage{caption}
\usepackage{graphicx}
\usepackage{subfigure}

\ifCLASSINFOpdf
\else
\fi
\usepackage{amsmath}
\begin{document}
\captionsetup{font={small}}
\title{Aggregated Feasible Region of Heterogeneous Demand-Side Flexible Resources---Part I: Theoretical Derivation of the Exact Model}
\author{Yilin~Wen,~\IEEEmembership{Student Member,~IEEE,}
        Zechun~Hu,~\IEEEmembership{Senior Member,~IEEE,}\\
        Shi~You,~\IEEEmembership{Member,~IEEE,}
        and~Xiaoyu~Duan,~\IEEEmembership{Student Member,~IEEE}

\thanks{This work was supported by xxx . \emph{(Corresponding author: Zechun Hu.)}}
\thanks{Yilin Wen, Zechun Hu and Xiaoyu Duan are with the Department of Electrical Engineering, Tsinghua University, Beijing 100084, China
(e-mail: wen-yl20@mails.tsinghua.edu.cn; zechhu@tsinghua.edu.cn; duanxy16@mails.tsinghua.edu.cn).}
\thanks{Shi You is with the Department of Electrical Engineering, Technical University of Denmark, Lyngby 2800, Denmark (e-mail: sy@elektro.dtu.dk).}
}

\maketitle

\begin{abstract}
  In the first part of the two-part series, the model to describe the exact aggregated feasible region (AFR) of multiple types of demand-side resources are derived. Based on a discrete-time unified individual model of heterogeneous resources, the calculation of AFR is, in fact, a feasible region projection problem. Therefore, the Fourier-Motzkin Elimination (FME) method is used for derivation. By analyzing the redundancy of all possible constraints in the FME process, the mathematical expression and calculation method for the exact AFR is proposed. The number of constraints is linear with the number of resources, and is exponential with the number of time intervals, respectively. The computational complexity has been dramatically simplified compared with the original FME. However, the number of constraints in the model is still exponential and cannot be simplified any more. Hence, In Part II of this paper, several approximation methods are proposed and analyzed in detail.
\end{abstract}
\IEEEpeerreviewmaketitle

\section{Introduction}
\IEEEPARstart{T}{he} high penetration of intermittent renewable energy sources such as wind power (WP) and photovoltaics (PVs) has increased the demand for grid regulation services. In this context, the power systems can no longer rely solely on the traditional units, and the demand-side flexible resources (DSFRs) will also play an essential role \cite{ma2013demand,chau2017demand}. Typical DSFRs include distributed generation (DG), electric vehicles (EVs), distributed energy storage (DES), and thermostatically controlled loads (TCLs). The regulation service provided by DSFRs, which mainly includes demand response and ancillary services, can increase the safety and stability of the power grid, promote the consumption of renewable energy, and reduce user electricity costs. Driven by the target of carbon peak and neutrality, China’s electricity market reform has encouraged the interaction between supply and demand \cite{liu2021optimal}. Time-of-use (TOU) electricity prices and load-participated ancillary service markets have been implemented in many places across China. The United States, Europe, and other places have also organized regulation services markets or demand response (DR) projects for DSFR. Price-based DR like TOU price is easy to participate and has been widely promoted \cite{huang2019demand}. However, since the control of price-based DR is indirect, it is usually not well-performed. Ancillary services and incentive-cased DR allow the system operator to dispatch DSFRs directly, resulting in better performance \cite{yilmaz2020analysis}, but they usually have requirements on the scale of the service provider. For instance, the PJM market requires that the adjustable power is not less than 0.1 MW \cite{pjmrule}. There, small-scale distributed demand-side resources should be aggregated as aggregators to reach the market entry threshold.

In the previous research, there are two main categories of modeling methods for DSFR scheduling. The first one is the optimization model \cite{6575202,DISOMMA2018272,9186055,6313962,7122924,9484074,9393595,9460799,9078044} that considers the flexibilities of all resources. Such model can be solved by centralized or distributed optimization algorithms. Although the centralized optimization method can theoretically find the global optimal solution, it has problems with scalability and privacy protection. Existing studies using the centralized optimization \cite{6575202,DISOMMA2018272,9186055,9460799,9078044} limit the scope of application to small-scale resources. Distributed optimization \cite{6313962,7122924,9484074,9393595} can be implemented with limited data exchange. Although it avoids the disadvantages of the centralized method, it brings new problems. The distributed optimization method requires each unit to have calculation and communication capabilities and typically has poor convergence and long calculation time in actual application. Therefore, it is hard to use for real-time computation. 

The second category is hierarchical dispatch \cite{7393621,7857788,7907339,9143177} in which multiple DSFRs are organized as aggregators. By this way, aggregators should submit their feasible regions of the total power to the power system operator, and then the system operator runs the market clearing or dispatching program. After that, the total power of each aggregator is determined, and the aggregator allocates the power to each resource for execution. Both \cite{7393621} and \cite{7857788} design a two-level hierarchical structure to dispatch various DSFRs. In \cite{7907339}, a hierarchical framework is proposed to dispatch EVs to hedge against wind power uncertainty. And in\cite{9143177}, the authors consider the hierarchical control of TCLs. Most of the current market clearing or scheduling rules about DSFR belong to this hierarchical structure. 

A vital issue of the hierarchical structure lies in formulating the flexibility of aggregated resources, that is, the feasible region of aggregated power, namely aggregated feasible region (AFR). The AFR should be carefully formed to ensure the lower-level power allocation can be successfully conducted. However, the model of AFR in current practical applications is usually determined by experience. For example, PJM only considers the upper and lower boundaries of total adjustable power \cite{pjmrule}. More precisely, the rule of the North China peak shaving market divides the DSFRs into two types. The first has power and energy capacity, including DES and vehicle-to-grid (V2G) charging piles. The second only has power capacity, including ordinary charging piles and TCLs\cite{huabeirule}. The AFRs described by these methods are generally far from the actual AFR.

Some literature discusses AFR specifically. In \cite{6832599}, the authors analyze the AFR of TCLs and point out that exact AFR is the Minkowski sum of the feasible region of each TCL equipment. Since this sum is difficult to calculate, the authors did not give the mathematical expression of the exact AFR but use a generalized electricity storage model to bound it. Besides, the method cannot be extended to other types of DSFR. In \cite{7393621}, a unified flexibility model of heterogeneous DSFRs is proposed, where the flexibility of each resource is described by the power and energy boundaries. The AFR is also formulated by these boundaries, and the parameters are the summation of parameters in the individual model. However, this AFR is also inaccurate. The state-space model is used to describe the flexibility of the EV aggregator in \cite{9042884}, but the state variables should be the energy of all EVs that is inconsistent with the purpose of aggregation.

As we know, there is little research that explicitly discusses the AFR, especially considering multiple types DSFRs. The existing models of AFR are inaccurate, leading to the problem of infeasible allocation in large-scale resource aggregation. Therefore, this paper mainly focuses on modeling the exact AFR of various types of DSFRs. In the first part of this two-series paper, we derive the expression of exact AFR based on the Fourier-Motzkin elimination (FME) \cite{SchrijverTheory} method, provide the calculation , and anayzed the computational complexity.

The rest of Part I of this paper is organized as follows. Section II introduces the theoretical basis. Section III-V discusses all the possibilities of the FME algorithm and gives the general formula of the AFR. Section VI gives the calculation method of the exact AFR and analyzes its computational complexity. Section VII concludes the paper.

\section{Theoretical basis}
\subsection{Unified Flexibility Model of Individual DSFR}
Typical DSFRs can be divided into three categories, namely generator-like resources and storage-like resources. The flexibility of generator-like resources is that the power is adjustable within a specific range. DGs belong to this category. The summation of all power boundaries can accurately describe the AFR of such resources. The storage-like resources, e.g. DESs, EVs, and TCLs are constrained by both the power and energy boundaries. The double constraints of power and energy makes the calculation of AFR of multiple DSFRs extremely complicated. Therefore, this paper mainly discusses the storage-like resources.

After discretizing the time according to the interval of $\Delta t$, the operating constraints of a storage-like DSFR $i$ can be expressed in terms of power and energy boundaries by (\ref{individualcons}) \cite{7393621}, where, $e_i(t)$ are the energy variable of resource $i$ in interval $t$, $(\underline {{p_i}}(t) ,\overline {{p_i}}(t) ,\underline {{e_i}}(t) ,\overline {{e_i}}(t) )$ are the lower and upper boudnaries of power and energy, respectively. $\mathbb{C}_i^p(t)$ and $\mathbb{C}_i^e(t)$ be the power and energy constraints of resource $i$ in interval $t$, respectively. And $\{\mathbb{C}_i^p(t), \mathbb{C}_i^e(t)\}$ is denoted as $\mathbb{C}_i(t)$. In the rest of this paper, for the sake of brevity, set $\Delta t=1$ so that $\Delta t$ will not appear again. In practical applications, it is only needed to multiply all the power-related parameters by $\Delta t$.
\begin{equation}
  \label{individualcons}
  \begin{array}{*{20}{c}}
    {\Delta T\underline {{p_i}} (t) \le {e_i}(t) - {e_i}(t - 1) \le \Delta T\overline {{p_i}} (t):\mathbb{C}_i^p(t)}\\
    {\underline {{e_i}} (t) \le {e_i}(t) \le \overline {{e_i}} (t):\mathbb{C}_i^e(t)}
    \end{array},\forall t
\end{equation}

\subsection{Hypotheses of the Unified Flexibility Model}
First, hypotheses (Hypo \ref{hyp1}-\ref{hyp3}) should be made on the parameters for the individual flexibility model. Apparently, Hypo \ref{hyp1} should hold. Hypo \ref{hyp2} and \ref{hyp3} contain the construction principle of energy boundaries: the upper and lower energy boundaries correspond to the fastest and slowest energy change trajectories. Even though all these hypotheses hold in the individual models constructed in \cite{7393621}, they are not clearly claimed. These hypotheses are crucial to the subsequent derivations.
\newtheorem{hypothesis}{Hypo}
\begin{hypothesis}
  \label{hyp1}
  $\left\{ {\begin{array}{*{20}{c}}
    {\underline {{p_i}} (t) \le \overline {{p_i}} (t)}\\
    {\underline {{e_i}} (t) \le \overline {{e_i}} (t)}
    \end{array}} \right.,\forall t.$
\end{hypothesis}

\begin{hypothesis}
  \label{hyp2}
  $\left\{ {\begin{array}{*{20}{c}}
    {\underline {{p_i}} (t) \le \underline {{e_i}} (t) - \underline {{e_i}} (t - 1) \le \overline {{p_i}} (t)}\\
    {\underline {{p_i}} (t) \le \overline {{e_i}} (t) - \overline {{e_i}} (t - 1) \le \overline {{p_i}} (t)}
    \end{array}} \right.,\forall t.$
\end{hypothesis}

\begin{hypothesis}
  \label{hyp3}
  $\left\{ {\begin{array}{*{20}{c}}
    {\underline {{e_i}} (t - 1) + \overline {{p_i}} (t) \le \overline {{e_i}} (t)}\\
    {\overline {{e_i}} (t - 1) + \underline {{p_i}} (t) \ge \underline {{e_i}} (t)}
    \end{array}} \right.,\forall t.$
\end{hypothesis}

\subsection{Fourier-Motzkin Elimination}
FME is a classic algorithm for calculating the projection of polytopes. Let $\mathbf{x}=[x_1,x_2,...,x_n]$ be the variables of n-dimensional space, and $P=\{\mathbf{x}|\mathbf{A}\mathbf{x} \leq \mathbf{b}\}$ be the polytope to be projected, where the number of rows of $\mathbf{A}$ is $M$. The projection $P'=\{\mathbf{x}'|\mathbf{A}'\mathbf{x}' \leq \mathbf{b}'\}$ of $P$ on $\mathbf{x}'= [x_1,x_2,...,x_{n-1}]$ means $\mathbf{A}'\mathbf{x}' \leq \mathbf{b}'$ and $\mathbf{A}\mathbf{x} \leq \mathbf{b}$ have the same solution on $\mathbf{x}'$. To eliminate $x_n$, FME first classifies all constraints according to the signs of coefficients of $x_n$ by (\ref{fme1}). Among them, $Z$ does not participate in the elimination. Rearrange to obtain (\ref{fme2}) so the variable $x_n$ can be eliminated to get (\ref{fme3}).
\begin{equation}
  \label{fme1}
  \begin{array}{l}
    N = \{ i \in M|{a_{in}} < 0\}, \\
    Z = \{ i \in M|{a_{in}} = 0\}, \\
    P = \{ i \in M|{a_{in}} > 0\} 
    \end{array}
\end{equation}
\begin{equation}
  \label{fme2}
  \left\{ \begin{array}{l}
    {x_n} \ge {b_i}/{a_{in}} - \sum\limits_{k \ne n} {{a_{ik}}/{a_{in}}{x_k}} ,i \in N\\
    {x_n} \le {b_i}/{a_{in}} - \sum\limits_{k \ne n} {{a_{ik}}/{a_{in}}{x_k}} ,i \in P
    \end{array} \right.
\end{equation}
$\forall i \in N,j \in P:$
\begin{equation}
  \label{fme3}
  \begin{array}{r}
    {b_i}/{a_{in}} - \sum\limits_{k \ne n} {{a_{ik}}/{a_{in}}{x_k}}  \le {b_j}/{a_{jn}} - \sum\limits_{l \ne n} {{a_{jl}}/{a_{jn}}{x_j}}
    \end{array}
\end{equation}

The core of FME is to combine all the inequalities about the variable to be eliminated. It is noted that the number of new constraints added after eliminating one variable is $N \cdot P$. If multiple variables are to be eliminated, the complexity of this algorithm increases at a double exponential level \cite{linearint}. 

\subsection{Using Fourier-Motzkin Elimination to Calculate the AFR}

Let $T$ be the total number of time intervals, then the feasible region of an individual DSFR is a $T$-dimensional polytope. Let $\mathcal{N} (|\mathcal{N}|=N)$ be the set of all DFSRs, $i\in\mathcal{N}$ be one DSFR, and $E(t) = \sum\nolimits_{i \in {\mathcal N}} {{e_i}(t)} $ be the total energy of aggregator, then the AFR is the Minkowski sum of the individual feasible regions. However, there is no efficient algorithm for calculating the Minkowski sum of multiple high-dimensional polytopes. Hence, we consider the problem from another aspect: all the individual constraints consititute a polytope in the $NT$-dimensional space, so the AFR is also the projection of the polytope on the $E(t)$ space. The FME method can be used to eliminate $NT$ variables and $4NT$ constraints. The FME method also has a high complexity. If the order of calculation is arbitrary and the process is not simplified, the computational complexity is $O({(2NT)^{{2^{NT}}}})$. However, since FME is clear and easy to analyze, it is used for derivation. Before defining the order of FME, a nomenclature should be defined first: ${(\cdot)_{\mathcal X}} = \left\{ {{(\cdot)_i}|i \in \mathcal X} \right\}$, where $(\cdot)$ can be $e$, $\mathbb{C}$, $\mathbb{C}^p$, and $\mathbb{C}^e$, and $\mathcal X$ is a set of DSFRs. Then, the order of FME is defined as follows:
\newtheorem{definition}{Definition}
\begin{definition}
  The FME starts from the last interval $T$, and eliminates ${e_{\mathcal N}}(T)$, ${e_{\mathcal N}}(T-1)$, ..., ${e_{\mathcal N}}(1)$ in turn. Eliminating all ${e_i}(t), i\in \mathcal N$ is called a step of FME.
\end{definition}
\newtheorem{notation}{Notation}

 According to this order, when ${e_{\mathcal N}}( t  + 1)$ is eliminated, all the expressions about variables obtained are linear combinations of $E(T)$, $E(T - 1)$,..., $E( t  + 1)$ and ${e_{{\mathcal X}}}( t )({{\mathcal X}} \subseteq {{\mathcal N}})$. Let $q=T-t$ in the whole paper for expression convenience. Let $\mathbf{E}(q) := [ E(T),E(T - 1), \cdots ,E( t  + 1)]^T$ and ${{\mathbf{e}}_{{\mathcal X}}}( t )$ is the column vector composed of all ${{e}_{{\mathcal X}}}( t )$. After eliminating ${e_{\mathcal N}}( t +1)$, all the resulting expressions can be written in the form of \eqref{generalform}, where $\mathbf{v}$ and $\mathbf{k}$ are row vectors.
\begin{equation}
  \label{generalform}
  \underline {{\gamma _{{\mathbf{v}},{\mathbf{k}},{{\mathcal X}}}}}  \le 
    {{\mathbf{v}}{\mathbf{E}}(q) + }
    {{\mathbf{k}}{{\mathbf{e}}_{{\mathcal X}}}( t )}
     \le \overline {{\gamma _{{\mathbf{v}},{\mathbf{k}},{{\mathcal X}}}}}:{\mathbb{F}_{{\mathbf{v}},{\mathbf{k}},{{\mathcal X}}}}( t )
\end{equation}
Therefore, when FME reaches the $ t $-th interval, there are three methods to eliminate ${e_i}( t )$:
\begin{itemize}[\IEEEsetlabelwidth{Z}]
  \item[i)] Eliminate through the combination of $\mathbb{C}_{\mathcal{N}}( t )$.
  \item[ii)] Eliminate by combining $\mathbb{C}_{\mathcal{N}}( t )$ and ${\mathbb{F}_{{\mathbf{v}},{\mathbf{k}},{{\mathcal X}}}}( t ),\forall {\mathbf{v}}, {\mathbf{k}}, \mathcal X$.
  \item[iii)] Eliminate through the combination of ${\mathbb{F}_{{\mathbf{v}},{\mathbf{k}},{{\mathcal X}}}}( t ),\forall {\mathbf{v}}, {\mathbf{k}}, \mathcal X$.
\end{itemize}

The following three sections will discuss these three methods separately.
\section{The First Method of Elimination}\label{SST1}
This method is the simplest, it can be further divided into two cases.
\subsection{Combine $\mathbb{C}_i^p( t )$ and $\mathbb{C}_i^e( t )$}\label{redundant0}
The result of this combination is (\ref{cnquchu}). However, due to Hypo \ref{hyp2}, inequalities (\ref{cnquchu}) are implied by $\mathbb{C}_i^e( t -1)$, which will be used in the next step. So the constraints obtained in this case are redundant.
  \begin{equation}
    \label{cnquchu}
    \underline {{e_i}} ( t ) - \overline {{p_i}} ( t ) \le {e_i}( t  - 1) \le \overline {{e_i}} ( t ) - \underline {{p_i}} ( t )
  \end{equation}
\subsection{Combine $\mathbb{C}_i( t )$ and $\mathbb{C}_{\mathcal N -  \{i\}}( t )$ to construct $E( t ) $}
In this case, there is no need to concern the specific $i$, and only need to pay attention to which DSFRs use $\mathbb{C}^p( t )$ and which DSFRs use $\mathbb{C}^e( t )$ to participate in the combination. Let $\mathcal X$ be the DSFR set that uses $\mathbb{C}^p( t )$, then ${{\mathcal N}- {\mathcal X}}$ is the set that uses $\mathbb{C}^e( t )$. The inequality obtained by this method is (\ref{cnbuqi}). 
\begin{equation}
  \label{cnbuqi}
  \begin{array}{l}
    \sum\limits_{i \in {{\mathcal N}-{\mathcal X}}} {\underline {{e_i}} ( t )}  + \sum\limits_{i \in {{\mathcal X}}} {\underline {{p_i}} ( t )}  \\
      \le E( t ) - \sum\limits_{i \in {{\mathcal X}}} {{e_i}( t -1)}  
  \\
      \le\sum\limits_{i \in {{\mathcal N}-{\mathcal X}}} {\overline {{e_i}} ( t )}  + \sum\limits_{i \in {{\mathcal X}}} {\overline {{p_i}} ( t )} 
\end{array} \forall \mathcal X \subseteq \mathcal N
\end{equation}

This constraint is not redundant. If $\mathcal{X} \neq \varnothing $ or $\mathcal N$, it should participate in the following elemination steps. In particular, when $\mathcal{X} = \varnothing $, formula (\ref{cnbuqi}) is $\sum_{i \in \mathcal N} {\underline {{e_i}} ( t )}  \le E( t ) \le \sum_{i \in \mathcal N} {\overline {{e_i}} ( t )} $, and when $\mathcal{X} = \mathcal{N} $, formula (\ref{cnbuqi}) is $\sum_{i \in \mathcal N} {\underline {{p_i}} ( t )}  \le E( t ) - E( t  - 1) \le \sum_{i \in \mathcal N} {\overline {{p_i}} ( t )} $, which are the aggregated power and energy constraints.

\section{The Second Method of Elimination}\label{SST2}
\subsection{Feature Analysis and Operation Definition}
In the previous section, set $t=T$ to get the result of the first step of elimination. Except for $\mathcal{X} = \varnothing\;\text{or}\;\mathcal{N} $, all other inequalities will be used in the next step. Note that the coefficients of $e_{\mathcal{X}}( t )$ in \eqref{cnbuqi} are either all $1$ or all $-1$, meaning that the coefficient vector $\mathbf{k}$ in expression \eqref{generalform} is $\mathbf{1}$ or $\mathbf{-1}$. This feature is used as an induction hypothesis in this section, which will be proved maintained in the second method. In this section, the $e_{\mathcal{X}}( t )$ part in the ${\mathbb{F}_{{\mathbf{v}},{\mathbf{k}},{{\mathcal X}}}}( t )$ expression is mainly discussed, so $b \le \sum_{i \in {{\mathcal X}}} {{e_{i}}( t )}  \le a$ is used to represent ${\mathbb{F}_{{\mathbf{v}},{\mathbf{k}},{{\mathcal X}}}}( t )$, where $\mathbf{E}(t)$ is included in $a$ and $b$. The expression in \eqref{generalform} will be reused when needed. All the operations of combining $b \le \sum_{i \in {{\mathcal X}}} {{e_{i}}( t )}\le a$ and $\mathbb{C}_{\mathcal{N}}( t )$ are defined as follows:

\begin{definition}
  Combining $\mathbb{C}_{\mathcal{N}-\mathcal{X}}( t )$ and $b \le \sum_{i \in {{\mathcal X}}} {{e_{i}}( t )}\le a$ to construct $\sum\limits_{i \in {{\mathcal N}}} {{e_i}( t )}  = E( t )$ and to construct $\sum\limits_{i \in {{\mathcal Y}}} {{e_i}( t )} (\mathcal Y\supsetneq \mathcal X)$ are called \emph{supplement all (SA)} and \emph{supplement part (SP)}, respectively.
\end{definition}

\begin{definition}
  Combining $\mathbb{C}_{\mathcal{X}}( t )$ and $b \le \sum_{i \in {{\mathcal X}}} {{e_{i}}( t )}\le a$ to erase $\sum_{i\in\mathcal{X}}e_i( t )$ is called \emph{remove all (RA)}, and to remove part of elements in $\mathcal{X}$ to get $\sum_{i\in\mathcal{Y}}e_i( t ) (\mathcal Y\subsetneq \mathcal X)$ is called \emph{remove part (RP)}.
\end{definition}

Once the SA or RA operation is done for a constraint, $e_{\mathcal{N}}( t )$ is no longer included in the resulting inequality, then this step of elimination is completed. It is not difficult to verify that the coefficients of $e_i( t )$ in the resulting inequalities obtained by SA or RA are still either all $1$ or all $-1$. Nevertheless, in addition to SA and RA, there are some other methods to conduct the elimination: do SP first and then do RA, namely \emph{SP-RA}, or do RP first and then do SA, namely \emph{RP-SA}, or loop SP and RP several times and finish by SA or RA, e.g. SP-RP-SP-RA, namely \emph{loop-SR}. Theorem \ref{SPRAredundancy} indicates that the inequalities obtained by SP-RA and RP-SA are redundant. For loop-SR, this theorem can be used several times to prove the implication relationship from the end to the start. For example, SP-RP-SP-RA is implied by SP-RP-RA, which is equal to SP-RA, implied by RA. In other words, since the chain of loop-SR must end with SA or RA, no matter how many switches of SP and RP are included in the chain, it is implied by a single SA or RA. Moreover, if it ends with SA, then the entire chain will be implied by SA, and RA is the same. Therefore, all the above operations except single SA or RA are redundant.

\newtheorem{theorem}{Theorem}
\begin{theorem}\label{SPRAredundancy}
  For the inequality $b \le \sum_{i \in {{\mathcal X}}} {{e_{i}}( t )}  \le a$, do SP first to get $\sum_{i \in {{\mathcal Y}}} {{e_{i}}( t )}$, and then do RA, the result is implied by doing RA directly for $b \le \sum_{i \in {{\mathcal X}}} {{e_{i}}( t )}  \le a$. Similarly, do RP for $b \le \sum_{i \in {{\mathcal X}}} {{e_{i}}( t )}  \le a$ first to get $\sum_{i \in {{\mathcal Z}}} {{e_{i}}( t )}$, and then do SA, the result is implied by doing SA directly for $b \le \sum_{i \in {{\mathcal X}}} {{e_{i}}( t )}  \le a$.
\end{theorem}

The first half of this theorem is proved in Appendix \ref{prfthm1}. The proof of the second half is very similar, so this paper omitted it. It is pointed out that the proof of this theorem is not trivial as it may look like because all the SP, SA, RP, and RA operations can choose either $\mathbb{C}^p( t )$ or $\mathbb{C}^e( t )$ selectively, and both Hypo \ref{hyp1} and \ref{hyp2} are used in the proof process. 
\subsection{Properties of the SA and RA operations}
When doing SA or RA, the constraints obtained are different depending on the selection of the DSFR sets that use $\mathbb{C}^p( t )$ and $\mathbb{C}^e( t )$. The specific forms of the results obtained by doing SA and RA for $\underline \gamma\le {{\mathbf{v}}}{\mathbf{E}}(q) \pm \sum_{i \in {{\mathcal X}}} {{e_i}( t )}  \le \overline \gamma$ are listed in Appendix \ref{OPRresults}, which can be regarded as a recursion formula. The following remark summarizes the properties of the results of SA and RA.

\newtheorem{remark}{Remark}
\begin{remark}
  \label{SARAsign}
  The SA operation produces $E( t )$ with the same sign as $e_{\mathcal X}( t )$, and $e_{\mathcal Y}( t -1) (\mathcal Y \subseteq \mathcal N -\mathcal X)$ with the opposite sign to $e_{\mathcal X}( t )$; the RA operation does not produce $E( t )$ but $e_{\mathcal Y}( t -1) (\mathcal Y \subseteq \mathcal X)$ with the same sign as $e_{\mathcal X}( t )$.
\end{remark}

According to this remark, it is easy to get the regularity of the coefficient of $E(t)$ in the second method, which will play an essential role in the AFR expression given later. However, the SA and RA operations need to be further simplified because one step of elimination will produce many constraints since the set $\mathcal Y$ is selected in all subsets of $\mathcal N -\mathcal X$  (for SA) or $\mathcal X$ (for RA), and $\mathcal X$ is selected in all non-trivial subsets of $\mathcal N$. Of course there are many redundant constraints, because $\mathcal Y$ may be generated from a different $\mathcal X$. Fortunately, these redundant constraints can be found out without computation. To this end, the following definitions are made first.
\begin{definition}
  \label{SAXYt}
  Among all the SA operations on $b \le \sum_{i \in {{\mathcal X}}} {{e_{i}}( t )}  \le a$, the operation that uses  $\mathbb{C}^p( t )$ for the set $\mathcal{Y} (\mathcal{Y} \subseteq \mathcal{N} -\mathcal{X} )$ and $\mathbb{C}^e( t )$ for the set $(\mathcal{N} -\mathcal{X})-\mathcal{Y}$ is called $\text{SA}_{\mathcal{X},\mathcal{Y}}( t )$. 
\end{definition}
\begin{definition}
  \label{RAXYt}
  Among all the RA operations on $b \le \sum_{i \in {{\mathcal X}}} {{e_{i}}( t )}  \le a$, the operation that uses  $\mathbb{C}^p( t )$ for the set $\mathcal{Y} (\mathcal{Y} \subseteq \mathcal{X} )$ and $\mathbb{C}^e( t )$ for the set $\mathcal{X}-\mathcal{Y}$ is called $\text{RA}_{\mathcal{X},\mathcal{Y}}( t )$. 
\end{definition}

Particularly, for $\mathcal X=\varnothing$ and $\mathcal X=\mathcal N$, $\text{SA}_{\varnothing,\mathcal{Y}}( t )$, $\text{SA}_{\mathcal N,\mathcal{Y}}( t )$, $\text{RA}_{\varnothing,\mathcal{Y}}( t )$ and $\text{RA}_{\mathcal N,\mathcal{Y}}( t )$ means to do nothing. The following theorem reveals the implication relationship of the operation RA and SA.
\begin{theorem}
  \label{OPRredundancy}
  For inequalities $b \le \sum_{i \in {{\mathcal X}}} {{e_i}( t )}  \le a$ and $\mathcal Y \neq \varnothing$, the following propositions hold:
  \begin{itemize}[\IEEEsetlabelwidth{Z}]
    \item[a)] Doing $\text{SA}_{\mathcal{X},\mathcal{Y}}( t )$ and then doing $\text{SA}_{\mathcal{Y},\mathcal{Z}}( t -1)$ is implied by doing $\text{SA}_{\mathcal{X},{{{\mathcal N}} - ({{\mathcal X}} \cup {{\mathcal Z}})}}( t )$ first, then combining the result with $\mathbb{C}_{\mathcal X - \mathcal Z}^e( t -1)$, and finally doing $\text{SA}_{\mathcal{N}-\mathcal{Z},\mathcal{Z}}( t -1)$.
    \item[b)] Doing $\text{SA}_{\mathcal{X},\mathcal{Y}}( t )$ and then doing $\text{RA}_{\mathcal{Y},\mathcal{Z}}( t -1)$, is implied by doing $\text{SA}_{\mathcal{X},\mathcal{Z}}( t )$ first and then doing $\text{RA}_{\mathcal{Z},\mathcal{Z}}( t -1)$.
    \item[c)] Doing $\text{RA}_{\mathcal{X},\mathcal{Y}}( t )$ and then doing $\text{SA}_{\mathcal{Y},\mathcal{Z}}( t -1)$ is implied by doing $\text{RA}_{\mathcal{X},\mathcal{X}\cap(\mathcal{N}-\mathcal{Z})}( t )$ first, then combing the result with $\mathbb{C}_{(\mathcal{N} -\mathcal{X}) - \mathcal Z}^e( t -1)$, and finally doing $\text{SA}_{\mathcal{N}-\mathcal{Z},\mathcal{Z}}( t -1)$.
    \item[d)] Doing $\text{RA}_{\mathcal{X},\mathcal{Y}}( t )$ and then doing $\text{RA}_{\mathcal{Y},\mathcal{Z}}( t -1)$, the result will be implied by doing $\text{RA}_{\mathcal{X},\mathcal{Z}}( t )$ first and then doing $\text{RA}_{\mathcal{Z},\mathcal{Z}}( t -1)$.
  \end{itemize}
\end{theorem}

\begin{figure}[!t]
  \centering
  \includegraphics[width=3.4in]{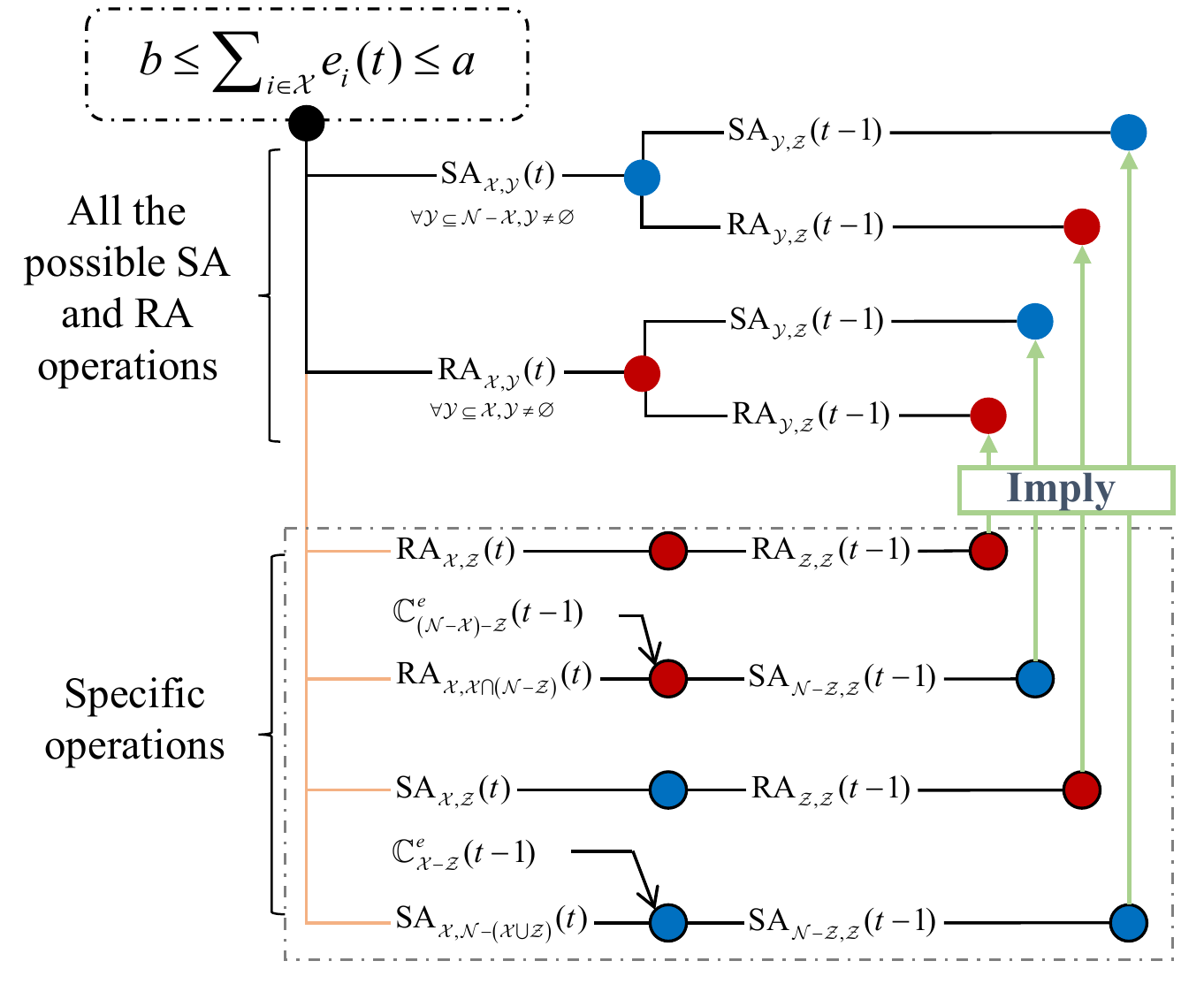}
  \caption{The implication relationship explained by Theorem \ref{OPRredundancy}: all the inequalities that RA and SA may generate are implied by a few specific inequalities.}
  \label{implyrelation}
\end{figure}

Theorem \ref{OPRredundancy} is proved in Appendix \ref{prfthm2}, where both Hypo \ref{hyp2} and \ref{hyp3} are used. The implication relationship in the theorem is illustrated in Fig. \ref{implyrelation}. Theorem \ref{OPRredundancy} reveals that all the inequalities that RA and SA may generate will be implied by a few specific inequalities. Specifically, the intermediate set $\mathcal Y$ is no longer selected in all subsets of $\mathcal N -\mathcal X$  (for SA) or $\mathcal X$ (for RA). Thus, a lot of redundant calculations can be avoided. For $\mathcal Z \neq \varnothing$, among all the inequalities about $\sum_{i \in {{\mathcal Z}}} {{e_i}( t -2)}$ generated by SA or RA, only the inequalities with the previous operation $\text{SA}_{\mathcal{N}-\mathcal{Z},\mathcal{Z}}( t -1)$ or $\text{RA}_{\mathcal{Z},\mathcal{Z}}( t -1)$ are not redundant. For $\mathcal Z =\varnothing$, in proposition a), it is equivalent to the case of $\mathcal Z =\mathcal X$; in proposition c), it is equivalent to the case of $\mathcal Z =\mathcal N - \mathcal X$; in propositions b) and d), it means that the operation $\text{RA}_{\mathcal{Z},\mathcal{Z}}( t -1)$ does nothing. Hence, the following corollary can be easily put forward without proof.
\newtheorem{corollary}{Corollary}
\begin{corollary}
  \label{legalOPR}
  Among all the operations for inequalities $b \le \sum_{i \in {{\mathcal X}}} {{e_i}( t )}  \le a$, only $\text{SA}_{\mathcal{X},\mathcal{N}-\mathcal{X}}( t )$, $\text{RA}_{\mathcal{X},\mathcal{X}}( t )$ and $\text{SA}_{\mathcal{X},\varnothing}( t )$ are not redundant.
\end{corollary}

The inequalities produced by $\text{SA}_{\mathcal{X},\mathcal{N}-\mathcal{X}}( t )$ and $\text{RA}_{\mathcal{X},\mathcal{X}}( t )$ will participate in the next step of elimination, and the inequalities produced by $\text{SA}_{\mathcal{X},\varnothing}( t )$ are only about $\mathbf{E}(q+1)$, which will be retained as members of the final constraints set of AFR.
\subsection{General Formula}\label{recform}

\begin{figure}[!t]
  \centering
  \includegraphics[width=3.4in]{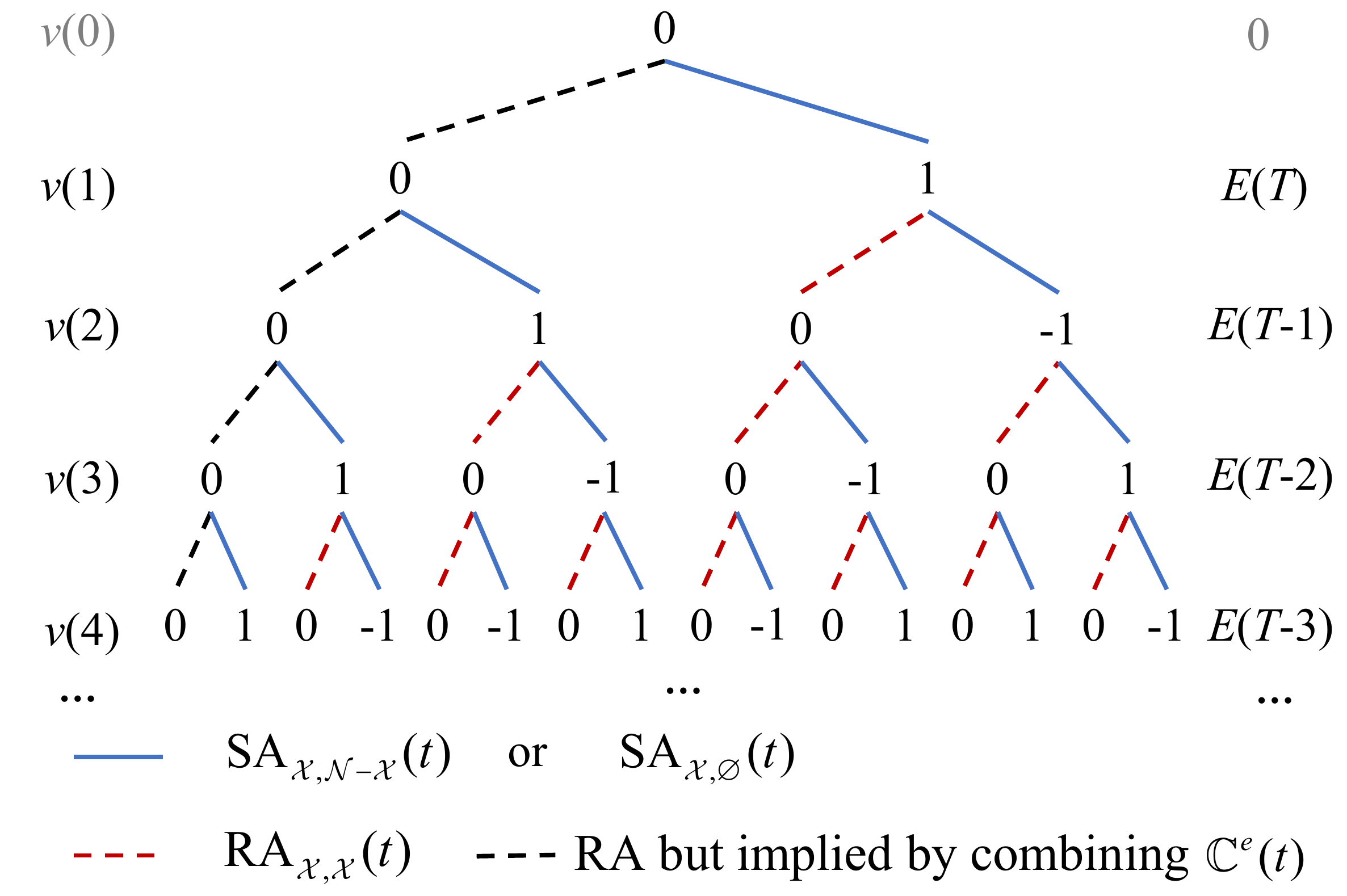}
  \caption{The development of the coefficient vector $\mathbf{v}$.}
  \label{coefE}
\end{figure}

Based on Corollary \ref{legalOPR}, the general formula of constraints produced in the second method of elimination can be derived through the Mathematical Induction method. In step $q$, the new $E( t )$ coefficient $v(q)$ satisfies the law given by Remark \ref{SARAsign}. Therefore, the development of coefficient $v(q)$ can be described by a complete binary tree, as illustrated in Fig. \ref{coefE}. In this binary tree, the right branch of a node represents the $\text{SA}_{\mathcal{X},\mathcal{N}-\mathcal{X}}( t )$ or $\text{SA}_{\mathcal{X},\varnothing}( t )$ operation, the left branch (except for the leftmost path) represents the $\text{RA}_{\mathcal{X},\mathcal{X}}( t )$ operation. The branches on leftmost path does not correspond to the $\text{RA}_{\mathcal{X},\mathcal{X}}( t )$ operation but to the combination of $\mathbb{C}^e( t )$ because of the redundant situation described in Section \ref{redundant0}. The paths from the root to the leaf have a one-to-one correspondence with values of the coefficient vector $\mathbf{v}$. 

Some nomenclature needs to be introduced before giving the the general formula. The set of generated paths when the elimination reaches the $q$-th step is denoted as $\mathcal L (q), l\in \mathcal L(q)$, and the vector $\mathbf{v}$ is denoted as $\mathbf{v}_l$ to indicate the corresponding relationship between $\mathbf{v}$ and $l$. Let ${\Phi _l}( q ) := {{\mathbf{v}}_l}{\mathbf{E}}(q)$, and ${{\mathbf{u}}_l}$ is generated by ${{\mathbf{v}}_l}$ according to \eqref{udef}. For example, given ${{\mathbf{v}}_l} = {[1,0, - 1,0,0,1]}$, then ${{\mathbf{u}}_l} = {[1,1,0,0,0,1]}$.  
\begin{equation}
  \label{udef}
  {u_l}(\theta) = \sum\limits_{\vartheta   = 1}^\theta {{v_l}(\vartheta )}, \forall \theta \in [1,q]
\end{equation}
Then ${\Phi _l}( q )$ can be defined in another form, that is ${\Phi _l}(q) := {\mathbf{u}}_l{\mathbf{P}}(q)$. The definition of ${\mathbf{P}}(q)$ is given by \eqref{Pdef}. Such a definition is convenient to interpret the physical meaning of ${\Phi _l}(q)$: since each component of $\mathbf{u}_l$ can be either $1$ or $0$, ${\Phi _l}(q)$ can be regarded as the summation of a certain combination of power in all intervals from $T$ to $ t +1$.
\begin{small}
\begin{equation}
  \label{Pdef}
  {\bf{P}}(q): = \left[ {\begin{array}{*{20}{c}}
    {P(T)}\\
    {P(T - 1)}\\
    {...}\\
    {P(t + 1)}
    \end{array}} \right] = \left[ {\begin{array}{*{20}{c}}
    {E(T) - E(T - 1)}\\
    {E(T - 1) - E(T - 2)}\\
    {...}\\
    {E(t + 1) - E(t)}
    \end{array}} \right]
\end{equation}
\end{small}
Finaly, let ${\Psi _{l,{{\mathcal X}}}}(q) := {\Phi _l}(q) + {I_l}\sum\limits_{i \in {{\mathcal X}}} {{e_i}( t )} $, where $I_l$ is the \emph{indicator function} of path $l$, defined as
\begin{small}
\begin{equation}
  \label{Idef}
  {I_l} = \left\{ {\begin{array}{*{20}{c}}
    {1,{\text{the number of non-zero nodes on }}l{\text{ is even}}},\\
    {- 1,{\text{the number of non-zero nodes on }}l{\text{ is odd}}}
    \end{array}} \right..\nonumber
\end{equation}
\end{small}
\begin{remark}
  $I_l=1\Leftrightarrow u_l(q)=1$, $I_l=-1\Leftrightarrow u_l(q)=0$. 
\end{remark}

The general formula of constraints is divided into two parts. The first part is \eqref{recformula_r}, corresponding to doing $\text{SA}_{\mathcal{X},\mathcal{N}-\mathcal{X}}$ and $\text{RA}_{\mathcal{X},\mathcal{X}}$ in all the previous elimination steps (from $T$ to $ t +1$), which will keep participating in the subsequent steps of elimination.
\\$\forall q \in [0,T],\forall l \in \mathcal L(q), \forall \mathcal X \subsetneq \mathcal N, \mathcal X \neq \varnothing$:
\begin{equation}
  \label{recformula_r}
  \underline {{\psi _{{{\mathcal X}},l}}} (q) \le {\Psi _{l,{{\mathcal X}}}}(q) \le \overline {{\psi _{{{\mathcal X}},l}}} (q),
\end{equation}
where 
\begin{small}
\begin{equation}
  \label{varphidefup}
  \begin{array}{l}
    \overline {{\psi _{{{\cal X}},l}}} (q) = \\
    \left\{ \begin{array}{l}

    {I_l} = 1:\\
    \sum\limits_{i \in {{\cal N}} - {{\cal X}}} {\left( {\overline {{e_i}} ({s_l}) - \left( {{\bf{1}} - {{\bf{u}}_l}_{[{g_l}:q - 1]}} \right){{\underline {{{\bf{p}}_i}} }_{[{g_l}:q - 1]}}} \right)} \\
    \;\;\;\;{ + \sum\limits_{i \in {{\cal X}}} {{{\bf{u}}_l}_{[{g_l}:q - 1]}{{\overline {{{\bf{p}}_i}} }_{[{g_l}:q - 1]}}} },\\
    {I_l} =  - 1:\\
    {\sum\limits_{i \in {{\cal X}}} {\left( {\overline {{e_i}} ({s_l}) - \left( {{\bf{1}} - {{\bf{u}}_l}_{[{g_l}:q - 1]}} \right){{\underline {{{\bf{p}}_i}} }_{[{g_l}:q - 1]}}} \right)} }\\
    \;\;\;\;{ + \sum\limits_{i \in {{\cal N}} - {{\cal X}}} {{{\bf{u}}_l}_{[{g_l}:q - 1]}{{\overline {{{\bf{p}}_i}} }_{[{g_l}:q - 1]}}} }
    
    \end{array} \right.
    \end{array},
\end{equation}

\begin{equation}
  \label{varphideflow}
  \begin{array}{l}
    \underline {{\psi _{{{\cal X}},l}}} (q) = \\
    \left\{ \begin{array}{l}
    {I_l} = 1:\\
    \sum\limits_{i \in {{\cal N}} - {{\cal X}}} {\left( {\underline {{e_i}} ({s_l}) - \left( {{\bf{1}} - {{\bf{u}}_l}_{[{g_l}:q - 1]}} \right){{\overline {{{\bf{p}}_i}} }_{[{g_l}:q - 1]}}} \right)} \\
    \;\;\;\; + \sum\limits_{i \in {{\cal X}}} {{{\bf{u}}_l}_{[{g_l}:q - 1]}{{\underline {{{\bf{p}}_i}} }_{[{g_l}:q - 1]}}} ,\\
    {I_l} =  - 1:\\
    \sum\limits_{i \in {{\cal X}}} {\left( {\underline {{e_i}} ({s_l}) - \left( {{\bf{1}} - {{\bf{u}}_l}_{[{g_l}:q - 1]}} \right){{\overline {{{\bf{p}}_i}} }_{[{g_l}:q - 1]}}} \right)} \\
    \;\;\;\;+ \sum\limits_{i \in {{\cal N}} - {{\cal X}}} {{{\bf{u}}_l}_{[{g_l}:q - 1]}{{\underline {{{\bf{p}}_i}} }_{[{g_l}:q - 1]}}} 
    \end{array} \right.
    \end{array},
\end{equation}
\end{small}
\begin{small}
\begin{equation}
  \underline {{{\bf{p}}_i}}  = \left[ \begin{array}{c}
    \underline {{p_i}} (T)\\
    \underline {{p_i}} (T - 1)\\
     \cdots \\
    \underline {{p_i}} (1)
    \end{array} \right],\overline {{{\bf{p}}_i}}  = \left[ \begin{array}{c}
    \overline {{p_i}} (T)\\
    \overline {{p_i}} (T - 1)\\
     \cdots \\
    \overline {{p_i}} (1)
    \end{array} \right],\nonumber
\end{equation}
\end{small}
the operator $[\cdot]_{[a:b]}$ selects the $a$-th to $b$-th elements of a vector, $\mathbf{1}$ is the row vector with all $1$ components and the length is equal to the required length, and $g_l$ is the index of the first non-zero element on path $l$, $s_l = T-g_l$. When $q=T$, the inequalities \eqref{recformula_r} contain the variable $e_i(0)$. By specify $e_i(0)=0,\forall i \in \mathcal N$ and $E(0)=0$, the FME will end at $q=T$.

The second part is \eqref{recformula_r1}, corresponding to the operation $\text{SA}_{\mathcal{X},\varnothing}$, which has no variable to eliminate and should be elements in the constraint set of AFR.\\
$\forall q \in [0,T],\forall l \in \mathcal L(q), \forall \mathcal X \subsetneq \mathcal N, \mathcal X \neq \varnothing$:
\begin{equation}
  \label{recformula_r1}
  \underline {{\phi _{{{\mathcal X}},l}}} (q) \le {\Phi _l}(q) \le \overline {{\phi _{{{\mathcal X}},l}}} (q),
\end{equation}
where
\begin{small} 
\begin{equation}\label{phiupdef}
  \begin{array}{l}
    \overline {{\phi _{{{\cal X}},l}}} (q) = \\
    \left\{ \begin{array}{l}
    {I_l} = 1:\\
    \begin{array}{c}
    {\sum\limits_{i \in {{\cal N}}- {{\cal X}}} {\left( {\overline {{e_i}} ({s_l}) - \left( {{\bf{1}} - {{\bf{u}}_l}_{[{g_l}:q - 1]}} \right){{\underline {{{\bf{p}}_i}} }_{[{g_l}:q - 1]}}} \right)} }\\
    { + \sum\limits_{i \in {{\cal X}}} {\left( {{{\bf{u}}_l}_{[{g_l}:q - 1]}{{\overline {{{\bf{p}}_i}} }_{[{g_l}:q - 1]}} + \overline {{e_i}} (t + 1)} \right)} }
    \end{array},\\
    {I_l} =  - 1:\\
    \begin{array}{l}
    {\sum\limits_{i \in {{\cal X}}} {\left( {\overline {{e_i}} ({s_l}) - \underline {{e_i}} (t + 1)\left. { - \left( {{\bf{1}} - {{\bf{u}}_l}_{[{g_l}:q - 1]}} \right){{\underline {{{\bf{p}}_i}} }_{[{g_l}:q]}}} \right)} \right.} }\\
    \;\;\;\;{ + \sum\limits_{i \in {{\cal N}}- {{\cal X}}} {{{\bf{u}}_l}_{[{g_l}:q]}{{\overline {{{\bf{p}}_i}} }_{[{g_l}:q]}}} }.
    \end{array}
    \end{array} \right.
    \end{array}
\end{equation}
\end{small}The definition of $\underline {{\phi _{{{\cal X}},l}}} (q)$ is symmetrical to $\overline {{\phi _{{{\cal X}},l}}} (q)$ with flipped upper and lower bounds, so it is omitted here.

\section{The Third Method of Elimination}\label{SST3}
In the previous section, the general formula of the constraints produced in the second method is derived, among which the constraints containing $e( t )$ can also be eliminated by combining each other, and that is the third method. The existence of this method becomes the fundamental reason why the overall computational complexity has risen to double exponential. However, the following theorem indicates that the general formula that we proposed in Section \ref{recform} represents all the effective inequalities in the FME process. Hence, the third method is unnecessary for computation.
\begin{theorem}\label{stt3redundancy}
  If ${{{\mathcal X}}_a} \cap {{{\mathcal X}}_b} \ne \varnothing $, for $j \in {{{\mathcal X}}_a} \cap {{{\mathcal X}}_b}$, all the constraints obtained by combining $\underline {{\psi _{{{{\mathcal X}}_a},{l_a}}}} (q) \le {\Psi _{{l_a},{{{\mathcal X}}_a}}}(q) \le \overline {{\psi _{{{{\mathcal X}}_a},{l_a}}}} (q),\forall {l_a} \in {{\mathcal L}}(q)$ and $\underline {{\psi _{{{{\mathcal X}}_b},{l_b}}}} (q) \le {\Psi _{{l_b},{{{\mathcal X}}_b}}}(q) \le \overline {{\psi _{{{{\mathcal X}}_b},{l_b}}}} (q),\forall {l_b} \in {{\mathcal L}}(q)$ to eliminate $e_j( t )$ are redundant.
\end{theorem}

This theorem is proved in Appendix \ref{prfthm3}, where Hypo \ref{hyp1} and \ref{hyp2} are used.

\section{Mathmatical Expression and Computational Complexity}

The derivation of Section \ref{SST1}-\ref{SST3} indicate that all the constraints in the AFR constraint set are \eqref{recformula_r1}, which are equivalent to
\\$\forall q\in [0,T], \forall l \in {{\mathcal L}}(q)$:
\begin{equation}\label{alg}
  \mathop {\max }\limits_{{{\mathcal Y}} \subsetneq {{\mathcal N}},{{\mathcal Y}} \ne \varnothing } \underline {{\phi _{{{\mathcal Y}},l}}} (q) \le {\Phi _l}(q) \le \mathop {\min }\limits_{{{\mathcal X}} \subsetneq {{\mathcal N}},{{\mathcal X}} \ne \varnothing  } \overline {{\phi _{{{\mathcal X}},l}}} (q).
\end{equation}

For a given $l \in \mathcal L (q)$, it seems that since $\mathcal X$ can be chosen arbitrarily from all the non-trivial subset of $\mathcal N$ so $2^N-2$ combinations need to be considered. However, it is pointed out that there is no need for so many calculations. Notice that the calculation methods of $\overline {{\phi _{{{\mathcal X}},l}}} (q)$ and $\underline {{\phi _{{{\mathcal X}},l}}} (q)$ can be regarded as the summation of two parts---$\mathcal X$ and $\mathcal N-\mathcal X$. A DSFR $i$ can be in either $\mathcal X$ or $\mathcal N-\mathcal X$, enabling the calculation of $\max$ and $\min$ to be decomposed to each DSFR. Specifically, $\overline {{\phi _{{{\mathcal X}},l}}} (q)$ can be calculated by the equation \eqref{minphiupdef}, and $\underline {{\phi _{{{\mathcal X}},l}}} (q)$ is symmetrical with it. There is no need to worry that $\overline {{\phi _{{{\mathcal X}},l}}} (q)$ may take the minimum value when $\mathcal X=\mathcal N$ or $\varnothing$ because $\mathcal X=\mathcal N$ or $\varnothing$ corresponds to the combination of inequalities \eqref{recformula_r1} itself with different $q$. Of course, the combined constraints are redundant, so it is impossible to get the minimum value of $\overline {{\phi _{{{\mathcal X}},l}}} (q)$ on $\mathcal X=\mathcal N$ or $\varnothing$.
\begin{small}
\begin{equation}\label{minphiupdef}
  \begin{array}{l}
    \mathop {\min }\limits_{{{\cal X}} \subsetneq {{\cal N}},{{\cal X}} \ne \varnothing } \overline {{\phi _{{{\cal X}},l}}} (q) = \mathop {\min }\limits_{{{\cal X}} \subseteq {{\cal N}}} \overline {{\phi _{{{\cal X}},l}}} (q) = \\
    \left\{ \begin{array}{l}
    {I_l} = 1:\\
    \sum\limits_{i \in {{\cal N}}} {\min \left\{ {\overline {{e_i}} ({s_l}) - \left( {{\bf{1}} - {{\bf{u}}_l}_{[{g_l}:q - 1]}} \right){{\underline {{{\bf{p}}_i}} }_{[{g_l}:q - 1]}}} \right.} \\
    \;\;\;\;\;\;\;\;\;\;\;\;\;\;\;\;\left. {,{{\bf{u}}_l}_{[{g_l}:q - 1]}{{\overline {{{\bf{p}}_i}} }_{[{g_l}:q - 1]}} + \overline {{e_i}} (t + 1)} \right\},\\
    {I_l} =  - 1:\\
    \sum\limits_{i \in {{\cal N}}} {\min \left\{ {\overline {{e_i}} ({s_l}) - \underline {{e_i}} (t + 1) - \left( {{\bf{1}} - {{\bf{u}}_l}_{[{g_l}:q - 1]}} \right){{\underline {{{\bf{p}}_i}} }_{[{g_l}:q - 1]}}} \right.} \\
    \;\;\;\;\;\;\;\;\;\;\;\;\;\;\;\;\left. {{\rm{,}}{{\bf{u}}_l}_{[{g_l}:q]}{{\overline {{{\bf{p}}_i}} }_{[{g_l}:q]}}} \right\}
    \end{array} \right.
    \end{array}
\end{equation}
\end{small}
In the outer layer ($\sum$ and $\min$), only $N$ times of two-number comparisons and one time of $N$-number summation are needed. In the inner layer, if sparse technology is used, the number of necessary calculations is related to the number of non-zero components on $\mathbf{u}_{l}$, so the overall time complexity of calculating $\mathop {\min }_{{{\mathcal X}} \subseteq {{\mathcal N}}} \overline {{\phi _{{{\mathcal X}},l}}} (q),\forall q \in [0,T],\forall l \in {\mathcal L}(q) $ is $\sum_{q = 0}^T {2N\left( \begin{array}{l}
  T\\  q  \end{array} \right)} =O(N{2^T})$, otherwise the complexity is $\sum_{{q} = 0}^T {Nq{2^{q}}} {\rm{ = }}O(NT{2^T})$.

It is emphasized that the number of total inequality constraints in the accurate AFR model is $2(2^T-1)$, which has nothing to do with $N$. Since formula \eqref{minphiupdef} calculates the sum of all DSFRs, it is straightforward to extend. To add a new DSFR, only need to add one item to the sum. In addition, to merge the AFRs of two DSFR sets, just directly add their $\overline {{\phi _{{{\mathcal X}},l}}} (q)$ and $\underline {{\phi _{{{\mathcal X}},l}}} (q)$ respectively to construct a larger AFR.

\section{Conclusion}
This paper uses the FME method to derive the exact AFR of heterogeneous demand-side flexible resources. The original FME method will produce $O({(2NT)^{{2^{NT}}}})$ constraints, many of which are redundant. Therefore, this paper analyzes the redundancy of all possible constraints generated by FME and finds the valid constraints through rigorous theoretical derivations and proofs. The mathematical expression and calculation method of the exact AFR is finally given. Compared with the original FME, the computational complexity is significantly reduced to $O(N2^T)$, where $N$ is the number of resources and $T$ is the number of time intervals. The number of constraints for the exact AFR is $2 (2^T-1)$, which has nothing to do with $N$, enabling the model conducive to expansion. However, if $T$ is large, e.g. $T\ge20$, the number of constraints is still extremely huge because it is exponential with $T$. In Part II of this paper, several approximation methods for the AFR are proposed to meet the needs of practical applications, and the effects of different models are discussed in detail.

\appendices
\section{Proof of Theorem 1}\label{prfthm1}
For the sake of brevity, only the proof about the inequality on the right side $\sum_{i \in {{\mathcal X}}} {{e_{i}}( t )}  \le a$ is provided here, and the proof of the left side is entirely symmetric.

The SP operation completes the original expression about $e_{\mathcal X}( t )$ to $e_{\mathcal Y}( t )$. Let $\mathcal{W}_p$ and $\mathcal{W}_e$ be the DSFR sets of which $\mathbb{C}^p( t )$ and $\mathbb{C}^e( t )$ are used in SP, respectively, and $\mathcal{V}_p$ and $\mathcal{V}_e$ be the DSFR sets of which $\mathbb{C}^p( t )$ and $\mathbb{C}^e( t )$ are used in RA, respectively. Apparently, the sets satisfy
\setcounter{equation}{0}
\renewcommand\theequation{A.\arabic{equation}}
\begin{small}
\begin{equation}\label{setsrelation}
  \begin{array}{c}
  \mathcal X\subseteq \mathcal Y,\\
  \mathcal W_p \cup \mathcal W_e = \mathcal Y - \mathcal X,\mathcal V_p \cup \mathcal V_e = \mathcal Y, \\
  \mathcal{W}_p\cup \mathcal{W}_e = \varnothing, \mathcal{V}_p\cup \mathcal{V}_e = \varnothing,\\
  \mathcal{W}_p-\mathcal{V}_p = \mathcal{W}_p\cap \mathcal{V}_e, \mathcal{W}_e-\mathcal{V}_p = \mathcal{W}_e\cap \mathcal{V}_e
  .\end{array}
\end{equation}
\end{small}
Doing SP-RA for the inequality gives that
\begin{small}
\begin{equation}
  \begin{array}{l}
    \sum\limits_{i \in {{{\mathcal V}}_p}} {{e_{i}}( t  - 1)}  - \sum\limits_{i \in {{{\mathcal W}}_p}} {{e_{i}}( t  - 1)}  \\
    \le a + \sum\limits_{i \in {{{\mathcal W}}_p}} {\overline {{p_i}} ( t )}  + \sum\limits_{i \in {{{\mathcal W}}_e}} {\overline {{e_i}} ( t )} 
  - \sum\limits_{i \in {{{\mathcal V}}_p}} {\underline {{p_i}} ( t )}  - \sum\limits_{i \in {{{\mathcal V}}_e}} {\underline {{e_i}} ( t )} 
  ,\end{array}\nonumber
\end{equation}
\end{small}
which is equivalent to
\begin{small}
\begin{equation}
  \label{SPRA}
  \begin{array}{l}
    \sum\limits_{i \in {{{\mathcal V}}_p} - {{{\mathcal W}}_p}} {{e_{i}}( t  - 1)}  - \sum\limits_{i \in {{{\mathcal W}}_p} - {{{\mathcal V}}_p}} {{e_{i}}( t  - 1)}   \\
    \le a + \sum\limits_{i \in {{{\mathcal W}}_p}} {\overline {{p_i}} ( t )}  + \sum\limits_{i \in {{{\mathcal W}}_e}} {\overline {{e_i}} ( t )} 
  - \sum\limits_{i \in {{{\mathcal V}}_p}} {\underline {{p_i}} ( t )}  - \sum\limits_{i \in {{{\mathcal V}}_e}} {\underline {{e_i}} ( t )} 
  .\end{array} 
\end{equation}
\end{small}Consider doing RA directly to the original inequality $\sum_{i \in {{\mathcal X}}} {{e_{i}}( t )}  \le a$, using $\mathbb{C}^p( t )$ for the set ${{{\mathcal V}}_p} - \left( {{{{\mathcal W}}_p} \cup {{{\mathcal W}}_e}} \right)$, and using $\mathbb{C}^e( t )$ for the set ${{{\mathcal V}}_e} - \left( {{{{\mathcal W}}_p} \cup {{{\mathcal W}}_e}} \right)$, we have
\begin{small}
\begin{equation}\label{SPRA1}
  \begin{array}{l}
    \sum\limits_{i \in {{{\mathcal V}}_p} - \left( {{{{\mathcal W}}_p} \cup {{{\mathcal W}}_e}} \right)} {{e_{i}}( t  - 1)}  \\
    \le a - \sum\limits_{i \in {{{\mathcal V}}_p} - \left( {{{{\mathcal W}}_p} \cup {{{\mathcal W}}_e}} \right)} {\underline {{p_i}} ( t )}  - \sum\limits_{i \in {{{\mathcal V}}_e} - \left( {{{{\mathcal W}}_p} \cup {{{\mathcal W}}_e}} \right)} {\underline {{e_i}} ( t )} 
    .\end{array}
\end{equation}
\end{small}
According to \eqref{setsrelation}, the following equalities hold:
\begin{small}
\begin{equation}
  \left\{ \begin{array}{l}
    {{{\mathcal V}}_p} - \left( {{{{\mathcal W}}_p} \cup {{{\mathcal W}}_e}} \right) = \left( {{{{\mathcal V}}_p} - {{{\mathcal W}}_p}} \right) - \left( {{{{\mathcal V}}_p} \cap {{{\mathcal W}}_e}} \right)\\
    {{{\mathcal V}}_e} - \left( {{{{\mathcal W}}_p} \cup {{{\mathcal W}}_e}} \right) = \left( {{{{\mathcal V}}_e} - {{{\mathcal W}}_p}} \right) - \left( {{{{\mathcal V}}_e} \cap {{{\mathcal W}}_e}} \right)
    .\end{array} \right.\nonumber
\end{equation}
\end{small}
Therefore, inequality \eqref{SPRA1} is equivalent to
\begin{small}
\begin{equation}
  \label{RA}
  \begin{array}{l}
    \sum\limits_{i \in {{{\mathcal V}}_p} - {{{\mathcal W}}_p}} {{e_{i}}( t  - 1)}  - \sum\limits_{i \in {{{\mathcal V}}_p} \cap {{{\mathcal W}}_e}} {{e_{i}}( t  - 1)}   \\
    \;\;\;\;\le a - \sum\limits_{i \in {{{\mathcal V}}_p} - {{{\mathcal W}}_p}} {\underline {{p_i}} ( t )}  + \sum\limits_{i \in {{{\mathcal V}}_p} \cap {{{\mathcal W}}_e}} {\underline {{p_i}} ( t )} \\
    \;\;\;\;\;\;\;\;- \sum\limits_{i \in {{{\mathcal V}}_e} - {{{\mathcal W}}_p}} {\underline {{e_i}} ( t )}  + \sum\limits_{i \in {{{\mathcal V}}_e} \cap {{{\mathcal W}}_e}} {\underline {{e_i}} ( t )} 
    .\end{array}
\end{equation}
\end{small}

To prove that \eqref{RA} implies \eqref{SPRA}, only need to prove
\begin{small}
\begin{equation}\label{onlyprove_a}
  \begin{array}{l}
    \sum\limits_{i \in {{{\mathcal V}}_p} \cap {{{\mathcal W}}_e}} {{e_{i}}( t  - 1)}  - \sum\limits_{i \in {{{\mathcal V}}_p} - {{{\mathcal W}}_p}} {\underline {{p_i}} ( t )} 
     + \sum\limits_{i \in {{{\mathcal V}}_p} \cap {{{\mathcal W}}_e}} {\underline {{p_i}} ( t )} \\
     \;\;\;\;\;\;\;\; - \sum\limits_{i \in {{{\mathcal V}}_e} - {{{\mathcal W}}_p}} {\underline {{e_i}} ( t )}  
     + \sum\limits_{i \in {{{\mathcal V}}_e} \cap {{{\mathcal W}}_e}} {\underline {{e_i}} ( t )}   \\
     \le \sum\limits_{i \in {{{\mathcal W}}_p} - {{{\mathcal V}}_p}} {{e_{i}}( t  - 1)}  + \sum\limits_{i \in {{{\mathcal W}}_p}} {\overline {{p_i}} ( t )}      + \sum\limits_{i \in {{{\mathcal W}}_e}} {\overline {{e_i}} ( t )} \\
     \;\;\;\;\;\;\;\; - \sum\limits_{i \in {{{\mathcal V}}_p}} {\underline {{p_i}} ( t )}  - \sum\limits_{i \in {{{\mathcal V}}_e}} {\underline {{e_i}} ( t )} 
    .\end{array}
\end{equation}
\end{small}
According to \eqref{setsrelation}, the following equalities hold:
\begin{small}
\begin{equation}
  \left\{ \begin{array}{l}
    {{{\mathcal V}}_p} - \left( {{{{\mathcal V}}_p} - {{{\mathcal W}}_p}} \right) = {{{\mathcal V}}_p} \cap {{{\mathcal W}}_p},\\
    {{{\mathcal V}}_e} - \left( {{{{\mathcal V}}_e} - {{{\mathcal W}}_p}} \right) = {{{\mathcal V}}_e} \cap {{{\mathcal W}}_p},\\
    {{{\mathcal W}}_p} = \left( {{{{\mathcal W}}_p} \cap {{{\mathcal V}}_p}} \right) \cup \left( {{{{\mathcal W}}_p} \cap {{{\mathcal V}}_e}} \right),\\
    {{{\mathcal W}}_e} = \left( {{{{\mathcal W}}_e} \cap {{{\mathcal V}}_p}} \right) \cup \left( {{{{\mathcal W}}_e} \cap {{{\mathcal V}}_e}} \right)
    .\end{array} \right.\nonumber
\end{equation}
\end{small}
Therefore, inequality \eqref{onlyprove_a} is equivalent to
\begin{small}
\begin{equation}\label{transed_a}
  \begin{array}{l}
    \sum\limits_{i \in {{{\mathcal V}}_p} \cap {{{\mathcal W}}_e}} {{e_{i}}( t  - 1)}  + \sum\limits_{i \in {{{\mathcal V}}_p} \cap {{{\mathcal W}}_p}} {\underline {{p_i}} ( t )}  + \sum\limits_{i \in {{{\mathcal V}}_p} \cap {{{\mathcal W}}_e}} {\underline {{p_i}} ( t )} \\
    \;\;\;\;\;\;\;\;
    + \sum\limits_{i \in {{{\mathcal V}}_e} \cap {{{\mathcal W}}_p}} {\underline {{e_i}} ( t )} 
    + \sum\limits_{i \in {{{\mathcal V}}_e} \cap {{{\mathcal W}}_e}} {\underline {{e_i}} ( t )}  \\
    \le \sum\limits_{i \in {{{\mathcal W}}_p} - {{{\mathcal V}}_p}} {{e_{i}}( t  - 1)}  + \sum\limits_{i \in {{{\mathcal V}}_p} \cap {{{\mathcal W}}_p}} {\overline {{p_i}} ( t )} 
     + \sum\limits_{i \in {{{\mathcal V}}_e} \cap {{{\mathcal W}}_p}} {\overline {{p_i}} ( t )} \\
     \;\;\;\;\;\;\;\; + \sum\limits_{i \in {{{\mathcal V}}_p} \cap {{{\mathcal W}}_e}} {\overline {{e_i}} ( t )}  
     +\sum\limits_{i \in {{{\mathcal V}}_e} \cap {{{\mathcal W}}_e}} {\overline {{e_i}} ( t )} 
    .\end{array}
\end{equation}
\end{small}According to Hypo \ref{hyp1}, if the following inequality holds, then \eqref{transed_a} holds.\
\begin{small}
\begin{equation}
  \begin{array}{l}
    \sum\limits_{i \in {{{\mathcal V}}_p} \cap {{{\mathcal W}}_e}} { {{e_{i}}} ( t  - 1)}  + \sum\limits_{i \in {{{\mathcal V}}_p} \cap {{{\mathcal W}}_e}} {\underline {{p_i}} ( t )}  
    + \sum\limits_{i \in {{{\mathcal V}}_e} \cap {{{\mathcal W}}_p}} {\underline {{e_i}} ( t )}   \\
    \;\;\;\;\le \sum\limits_{i \in {{{\mathcal W}}_p} - {{{\mathcal V}}_p}} { {{e_{i}}} ( t  - 1)}  + \sum\limits_{i \in {{{\mathcal V}}_e} \cap {{{\mathcal W}}_p}} {\overline {{p_i}} ( t )}  
    + \sum\limits_{i \in {{{\mathcal V}}_p} \cap {{{\mathcal W}}_e}} {\overline {{e_i}} ( t )} 
    \end{array}\nonumber
\end{equation}
\end{small}Since $\mathbb{C}_i^e( t -1):\underline {{e_{i}}} ( t  - 1) \le {e_{i}}( t  - 1) \le \overline {{e_{i}}} ( t  - 1)$ will be used in the subsequent elimination, it can be used as a known condition here. Hence, we only need to prove
\begin{small}
\begin{equation}\label{A_transb}
  \begin{array}{l}
    \sum\limits_{i \in {{{\mathcal V}}_p} \cap {{{\mathcal W}}_e}} {\overline {{e_{i}}} ( t  - 1)}  + \sum\limits_{i \in {{{\mathcal V}}_p} \cap {{{\mathcal W}}_e}} {\underline {{p_i}} ( t )}  
    + \sum\limits_{i \in {{{\mathcal V}}_e} \cap {{{\mathcal W}}_p}} {\underline {{e_i}} ( t )}   \\
    \;\;\;\;\le \sum\limits_{i \in {{{\mathcal W}}_p} - {{{\mathcal V}}_p}} {\underline {{e_{i}}} ( t  - 1)}  + \sum\limits_{i \in {{{\mathcal V}}_e} \cap {{{\mathcal W}}_p}} {\overline {{p_i}} ( t )}  
    + \sum\limits_{i \in {{{\mathcal V}}_p} \cap {{{\mathcal W}}_e}} {\overline {{e_i}} ( t )} 
    .\end{array}
\end{equation}
\end{small}Due to ${{{\mathcal W}}_p} - {{{\mathcal V}}_p} = {{{\mathcal V}}_e} \cap {{{\mathcal W}}_p}$ and Hypo \ref{hyp2}, if the following inequality holds, then \eqref{A_transb} holds.
\begin{small}
\begin{equation}\label{AtransC}
  \begin{array}{l}
    \sum\limits_{i \in {{{\mathcal V}}_p} \cap {{{\mathcal W}}_e}} {\overline {{e_{i}}} ( t )}  + \sum\limits_{i \in {{{\mathcal V}}_e} \cap {{{\mathcal W}}_p}} {\underline {{e_i}} ( t )}   \\
    \;\;\;\;\le \sum\limits_{i \in {{{\mathcal V}}_e} \cap {{{\mathcal W}}_p}} {\underline {{e_{i}}} ( t )}  + \sum\limits_{i \in {{{\mathcal V}}_p} \cap {{{\mathcal W}}_e}} {\overline {{e_i}} ( t )} 
    \end{array}
\end{equation}
\end{small}Inequality \eqref{AtransC} is equivalent to the constant inequality $0 \le 0$, so this completes the proof of Theorem \ref{SPRAredundancy}.

\section{}\label{OPRresults}
\setcounter{equation}{0}
\renewcommand\theequation{B.\arabic{equation}}
Doing SA for $\underline \gamma\le {{\mathbf{v}}}{\mathbf{E}}(q) + \sum_{i \in {{\mathcal X}}} {{e_i}( t )}  \le \overline \gamma$, all the inequalities to get are \eqref{SAresultpos}.
\begin{small}
\\$\forall \mathcal Y \subseteq \mathcal N - \mathcal X :$
\begin{equation}
  \label{SAresultpos}
  \begin{array}{l}
      \underline \gamma + \sum\limits_{i \in {{\mathcal Y}}} {\underline {{p_i}} ( t )}  + \sum\limits_{i \in \left( {{{\mathcal N}} - {{\mathcal X}}} \right) - {{\mathcal Y}}} {\underline {{e_i}} ( t )}  \\
      \;\;\;\;\le {\mathbf{v}}{\mathbf{E}}(q) + E( t ) - \sum\limits_{i \in {{\mathcal Y}}} {{e_i}( t  - 1)}  \\
      \;\;\;\;\;\;\;\;\le \overline \gamma  + \sum\limits_{i \in {{\mathcal Y}}} {\overline {{p_i}} } ( t ) + \sum\limits_{i \in \left( {{{\mathcal N}} - {{\mathcal X}}} \right) - {{\mathcal Y}}} {\overline {{e_i}} } ( t )
    \end{array}
\end{equation}
\end{small}
Doing SA for $\underline \gamma \le {{\mathbf{v}}}{\mathbf{E}}(q) - \sum_{i \in {{\mathcal X}}} {{e_i}( t )}  \le \overline \gamma$, all the inequalities to get are \eqref{SAresultneg}.\\
\begin{small}
$\forall \mathcal Y \subseteq \mathcal N - \mathcal X :$
\begin{equation}
  \label{SAresultneg}
  \begin{array}{l}
    \underline \gamma - \sum\limits_{i \in {{\mathcal Y}}} {\overline {{p_i}} } ( t ) - \sum\limits_{i \in \left( {{{\mathcal N}} - {{\mathcal X}}} \right) - {{\mathcal Y}}} {\overline {{e_i}} } ( t ) \\
    \;\;\;\;\le {\mathbf{v}}{\mathbf{E}}(q) - E( t ) + \sum\limits_{i \in {{\mathcal Y}}} {{e_i}( t  - 1)}  \\
    \;\;\;\;\;\;\;\;\le \overline \gamma - \sum\limits_{i \in {{\mathcal Y}}} {\underline {{p_i}} ( t )}  - \sum\limits_{i \in \left( {{{\mathcal N}} - {{\mathcal X}}} \right) - {{\mathcal Y}}} {\underline {{e_i}} ( t )} 
    \end{array}
\end{equation}
\end{small}
Doing RA for $\underline \gamma \le {{\mathbf{v}}}{\mathbf{E}}(q) + \sum_{i \in {{\mathcal X}}} {{e_i}( t )}  \le \overline \gamma$, all the inequalities to get are \eqref{RAresultpos}.\\
\begin{small}
$\forall \mathcal Y \subseteq \mathcal X :$
\begin{equation}
  \label{RAresultpos}
  \begin{array}{l}
    \underline \gamma - \sum\limits_{i \in {{\mathcal Y}}} {\overline {{p_i}} ( t )}  - \sum\limits_{i \in \left( {{{\mathcal N}} - {{\mathcal X}}} \right) - {{\mathcal Y}}} {\overline {{e_i}} ( t )}  \\
    \;\;\;\;\le {\mathbf{v}}{\mathbf{E}}(q) + \sum\limits_{i \in {{\mathcal Y}}} {{e_i}( t  - 1)}  \\
    \;\;\;\;\;\;\;\;\le \overline \gamma - \sum\limits_{i \in {{\mathcal Y}}} {\underline {{p_i}} ( t )}  - \sum\limits_{i \in \left( {{{\mathcal N}} - {{\mathcal X}}} \right) - {{\mathcal Y}}} {\underline {{e_i}} ( t )} 
    \end{array}
\end{equation}
\end{small}
Doing RA for $\underline \gamma\le {{\mathbf{v}}}{\mathbf{E}}(q) + \sum_{i \in {{\mathcal X}}} {{e_i}( t )}  \le \overline \gamma$, all the inequalities to get are \eqref{RAresultneg}.\\
\begin{small}
$\forall \mathcal Y \subseteq \mathcal X :$
\begin{equation}
  \label{RAresultneg}
  \begin{array}{l}
    \underline \gamma + \sum\limits_{i \in {{\mathcal Y}}} {\overline {{p_i}} ( t )}  + \sum\limits_{i \in \left( {{{\mathcal N}} - {{\mathcal X}}} \right) + {{\mathcal Y}}} {\overline {{e_i}} ( t )}  \\
    \;\;\;\;\le {\mathbf{v}}{\mathbf{E}}(q) - \sum\limits_{i \in {{\mathcal Y}}} {{e_i}( t  - 1)}  \\
    \;\;\;\;\;\;\;\;\le \overline \gamma + \sum\limits_{i \in {{\mathcal Y}}} {\underline {{p_i}} ( t )}  + \sum\limits_{i \in \left( {{{\mathcal N}} - {{\mathcal X}}} \right) + {{\mathcal Y}}} {\underline {{e_i}} ( t )} 
    \end{array}
\end{equation}
\end{small}

\section{Proof of Theorem 2}\label{prfthm2}
\setcounter{equation}{0}
\renewcommand\theequation{C.\arabic{equation}}
For the sake of brevity, only the proof about the right side inequality $\sum_{i \in {{\mathcal X}}} {{e_{i}}( t )}  \le a$ is provided here, and the proof of the left side is entirely symmetric. Only the detailed proofs of propositions a) and b) are given here, and that of c) and d) is omitted because they are very similar.
\subsection{Proof of Proposition a)}

Doing $\text{SA}_{\mathcal{X},\mathcal{Y}}( t ) (\mathcal{Y} \neq \varnothing, \mathcal{Y} \subseteq{\mathcal{N}-\mathcal{X}})$ and $\text{SA}_{\mathcal{Y},\mathcal{Z}}( t -1) (\mathcal{Z} \subseteq{\mathcal{N}-\mathcal{Y}})$ gives that
\begin{small}
\begin{equation}
  \label{SASAresult}
  \begin{array}{l}
    E( t ) - E( t  - 1) + \sum\limits_{i \in {{\mathcal Z}}} {{e_i}( t  - 2)}  \le 
    a + \sum\limits_{i \in {{\mathcal Y}}} {\overline {{p_i}} ( t )}  \\ 
    \;\;\;\;+\sum\limits_{i \in ({{\mathcal N}} - {{\mathcal X}}) - {{\mathcal Y}}} {\overline {{e_i}} ( t )}
     - \sum\limits_{i \in {{\mathcal Z}}} {\underline {{p_i}} ( t  - 1)}  - \sum\limits_{i \in ({{\mathcal N}} - {{\mathcal Y}}) - {{\mathcal Z}}} {\underline {{e_i}} ( t  - 1)} 
    .\end{array}
\end{equation}
\end{small}Consider doing $\text{SA}_{\mathcal{X},\mathcal{N}-(\mathcal{X}\cup\mathcal{Z})}( t )$, we get
\begin{small}
\begin{equation}
  \begin{array}{l}
    E( t ) - \sum\limits_{i \in {{\mathcal N}} - ({{\mathcal X}} \cup {{\mathcal Z}})} {{e_i}( t  - 1)} \\
    \;\;\;\;\le a + \sum\limits_{i \in {{\mathcal N}} - ({{\mathcal X}} \cup {{\mathcal Z}})} {\overline {{p_i}} ( t )}  + \sum\limits_{i \in {{\mathcal Z}} - {{\mathcal X}}} {\overline {{e_i}} ( t )} 
    .\end{array}
\end{equation}
\end{small}Combine the above inequality with $\mathbb{C}_{\mathcal X - \mathcal Z}^e( t -1)$, the result is
\begin{small}
\begin{equation}
\begin{array}{l}
  E( t ) - \sum\limits_{i \in {{\mathcal N}} - {{\mathcal Z}}} {{e_i}( t  - 1)} 
   \le a + \sum\limits_{i \in {{\mathcal N}} - ({{\mathcal X}} \cup {{\mathcal Z}})} {\overline {{p_i}} ( t )}     \\
   \;\;\;\;+ \sum\limits_{i \in {{\mathcal Z}} - {{\mathcal X}})} {\overline {{e_i}} ( t )}  - \sum\limits_{i \in {{\mathcal X}} \cup {{\mathcal Z}}} {\underline {{e_i}} ( t  - 1)} 
  .\end{array}
\end{equation}
\end{small}Doing $\text{SA}_{\mathcal{N}-\mathcal{Z},\mathcal{Z}}( t -1)$ for the above inequality gives that
\begin{small}
\begin{equation}\label{SACSAresult}
\begin{array}{l}
  E( t ) - E( t  - 1) + \sum\limits_{i \in {{\mathcal Z}}} {{e_i}( t  - 2)}  \le
  a + \sum\limits_{i \in {{\mathcal N}} - ({{\mathcal X}} \cup {{\mathcal Z}})} {\overline {{p_i}} ( t )}  \\
  \;\;\;\;+ \sum\limits_{i \in {{\mathcal Z}} - {{\mathcal X}}} {\overline {{e_i}} ( t )} - \sum\limits_{i \in {{\mathcal X}} \cup {{\mathcal Z}}} {\underline {{e_i}} ( t  - 1)}  - \sum\limits_{i \in {{\mathcal Z}}} {\underline {{p_i}} ( t  - 1)} 
  .\end{array}
\end{equation}
\end{small}To prove that \eqref{SACSAresult} implies \eqref{SASAresult}, just prove
\begin{small}
\begin{equation}
  \begin{array}{l}
    a + \sum\limits_{i \in {{\mathcal N}} - ({{\mathcal X}} \cup {{\mathcal Z}})} {\overline {{p_i}} ( t )}  + \sum\limits_{i \in {{\mathcal Z}} - {{\mathcal X}}} {\overline {{e_i}} ( t )} \\
    \;\;\;\;\;\;\;\;- \sum\limits_{i \in {{\mathcal X}} \cup {{\mathcal Z}}} {\underline {{e_i}} ( t  - 1)}  - \sum\limits_{i \in {{\mathcal Z}}} {\underline {{p_i}} ( t  - 1)}      \\
    \le a + \sum\limits_{i \in {{\mathcal Y}}} {\overline {{p_i}} ( t )}  + \sum\limits_{i \in ({{\mathcal N}} - {{\mathcal X}}) - {{\mathcal Y}}} {\overline {{e_i}} ( t )} \\
    \;\;\;\;\;\;\;\;- \sum\limits_{i \in {{\mathcal Z}}} {\underline {{p_i}} ( t  - 1)}  - \sum\limits_{i \in ({{\mathcal N}} - {{\mathcal Y}}) - {{\mathcal Z}}} {\underline {{e_i}} ( t  - 1)} 
    ,\end{array}
  \end{equation}
\end{small}which is equivalent to
\begin{small}
\begin{equation}
  \begin{array}{l}
    \sum\limits_{i \in ({{\mathcal N}} - {{\mathcal Y}}) - ({{\mathcal X}} \cup {{\mathcal Z}})} {\left( {\overline {{p_i}} ( t ) + \underline {{e_i}} ( t  - 1)} \right)} 
      \le \sum\limits_{i \in ({{\mathcal N}} - {{\mathcal Y}}) - ({{\mathcal X}} \cup {{\mathcal Z}})} {\overline {{e_i}} ( t )} 
    .\end{array}
  \end{equation}
\end{small}According to Hypo \ref{hyp3}, the above inequality holds. This complete the proof of proposition a).
\subsection{Proof of Proposition b)}
Doing $\text{SA}_{\mathcal{X},\mathcal{Y}}( t ) (\mathcal{Y} \neq \varnothing, \mathcal{Y} \subseteq{\mathcal{N}-\mathcal{X}})$ and $\text{RA}_{\mathcal{Y},\mathcal{Z}}( t -1) (\mathcal{Z} \subseteq{\mathcal{Y}})$ gives that
\begin{small}
\begin{equation}
  \label{SARAresult}
  \begin{array}{l}
    E( t ) - \sum\limits_{i \in {{\mathcal Z}}} {{e_i}( t  - 2)}  \le 
    a + \sum\limits_{i \in {{\mathcal Y}}} {\overline {{p_i}} ( t )}  + \sum\limits_{i \in ({{\mathcal N}} - {{\mathcal X}}) - {{\mathcal Y}}} {\overline {{e_i}} ( t )} 
      \\
      \;\;\;\;\;\;\;\;+ \sum\limits_{i \in {{\mathcal Z}}} {\overline {{p_i}} ( t  - 1)}  + \sum\limits_{i \in {{\mathcal Y}} - {{\mathcal Z}}} {\overline {{e_i}} ( t  - 1)} 
    .\end{array}
\end{equation}
\end{small}Consider doing $\text{SA}_{\mathcal{X},\mathcal{Z}}( t )$ first and then doing $\text{RA}_{\mathcal{Z},\mathcal{Z}}( t -1)$, we get
\begin{small}
\begin{equation}
  \label{SARAresult1}
  \begin{array}{l}
    E( t ) - \sum\limits_{i \in {{\mathcal Z}}} {{e_i}( t  - 2)}  \le 
    a + \sum\limits_{i \in {{\mathcal Z}}} {\overline {{p_i}} ( t )}  \\
    \;\;\;\;\;\;\;\;+ \sum\limits_{i \in ({{\mathcal N}} - {{\mathcal X}}) - {{\mathcal Z}}} {\overline {{e_i}} ( t )} 
     + \sum\limits_{i \in {{\mathcal Z}}} {\overline {{p_i}} ( t  - 1)} 
    .\end{array}
\end{equation}
\end{small}Comparing the right term of \eqref{SARAresult} and \eqref{SARAresult1}, if the following inequality holds, then \eqref{SARAresult1} implies \eqref{SARAresult}.
\begin{small}
\begin{equation}
  \label{Bfinal}
  \sum\limits_{i \in {{\cal Y}} - {{\cal Z}}} {\overline {{e_i}} ( t )}  \le \sum\limits_{i \in {{\cal Y}} - {{\cal Z}}} {\left( {\overline {{p_i}} ( t ) + \overline {{e_i}} ( t  - 1)} \right)} .
\end{equation}
\end{small}Inequality \eqref{Bfinal} holds due to Hypo \ref{hyp2}, so this complete the proof of proposition b).

\section{Proof of Theorem 3}\label{prfthm3}
\setcounter{equation}{0}
\renewcommand\theequation{D.\arabic{equation}}
Here we give the detailed proof when the numbers of non-zero nodes on $l_a$ and $l_b$ are both odd, that is, $I_{l_a}=I_{l_b}=1$. The other situations are similar, so only the key steps are given.

When $I_{l_a}=I_{l_b}=1$, ${e_j}( t )$ can be eliminated by combining ${\Psi _{{l_a},{{{\mathcal X}}_a}}}(q) \le \overline {{\psi _{{{{\mathcal X}}_a},{l_a}}}} (q)$ and ${\Psi _{{l_b},{{{\mathcal X}}_b}}}(q) \ge \underline {{\psi _{{{{\mathcal X}}_b},{l_b}}}} (q)$, or ${\Psi _{{l_b},{{{\mathcal X}}_b}}}(q) \le \overline {{\psi _{{{{\mathcal X}}_b},{l_b}}}} (q)$ and ${\Psi _{{l_a},{{{\mathcal X}}_a}}}(q) \ge \underline {{\psi _{{{{\mathcal X}}_a},{l_a}}}} (q)$. The two cases are entirely symmetrical, therefore, without loss of generality, we prove the previous one. In this case, the process of eliminating ${e_j}( t )$ is
\begin{small}
\begin{equation}
  \begin{array}{l}
    {\mathbf{u}}_{{l_a}}{\mathbf{P}}(q) - \sum\limits_{i \in {{{\mathcal X}}_a}- \{ j\} } {{e_i}( t )}  - \overline {{\psi _{{{{\mathcal X}}_a},{l_a}}}} (q) \\
    \;\;\;\;\le {e_j}( t ) \\
    \;\;\;\;\;\;\;\;\;\le {\mathbf{u}}_{{l_b}}{\mathbf{P}}(q) - \sum\limits_{i \in {{{\mathcal X}}_b}- \{ j\} } {{e_i}( t )}  - \underline {{\psi _{{{{\mathcal X}}_b},{l_b}}}} (q)
    .\end{array}
\end{equation}
\end{small}Eliminate ${e_j}( t )$ and rearrange to obtain
\begin{small}
\begin{equation}
  \label{stt3res}
  \begin{array}{l}
    ({\mathbf{u}}_{{l_a}} - {\mathbf{u}}_{{l_b}}){\mathbf{P}}(q) -\left( {\sum\limits_{i \in {{{\mathcal X}}_a}- \{ j\} } {{e_i}( t )}  - \sum\limits_{i \in {{{\mathcal X}}_b}- \{ j\} } {{e_i}( t )} } \right)\\
    \;\;\;\;\le \overline {{\psi _{{{{\mathcal X}}_a},{l_a}}}} (q) - \underline {{\psi _{{{{\mathcal X}}_b},{l_b}}}} (q)
    .\end{array}
\end{equation}
\end{small}It is noticed that in addition to the elimination of ${e_j}( t )$, all other variables of elements overlapping $\mathcal X _a$ and $\mathcal X _b$ are also eliminated. Let $\mathcal X$ be the set of all overlapping elements, ${{\mathcal X}} = {{{\mathcal X}}_a} \cap {{{\mathcal X}}_b}$. Define ${{\mathcal Y}} = {{{\mathcal X}}_a} - {{\mathcal X}}$, ${{\mathcal Z}} = {{{\mathcal X}}_b} - {{\mathcal X}}$, and ${{\mathcal W}} = {{\mathcal N}} - {{{\mathcal X}}_a} \cup {{{\mathcal X}}_b}$. It is obvious that the sets $\mathcal{X} $, $\mathcal{Y} $, $\mathcal{Z} $ and $\mathcal{W} $ do not intersect each other (the intersection is empty set). By this way, the inequality \eqref{stt3res} can be rewritten as \eqref{stt3res1}.
\begin{equation}
  \label{stt3res1}
  \begin{array}{l}
    ({\mathbf{u}}_{{l_a}} - {\mathbf{u}}_{{l_b}}){\mathbf{P}}(q) - \left( {\sum\limits_{i \in {{\mathcal Y}}} {{e_i}( t )}  - \sum\limits_{i \in {{\mathcal Z}}} {{e_i}( t )} } \right)\\
    \;\;\;\;\le \overline {{\psi _{{{\mathcal X}} \cup {{\mathcal Y}},{l_a}}}} (q) - \underline {{\psi _{{{\mathcal X}} \cup {{\mathcal Z}},{l_b}}}} (q)
    \end{array}
\end{equation}
According to \eqref{varphidefup} and \eqref{varphideflow}, $\overline {{\psi _{{{\cal X}} \cup {{\cal Y}},{l_a}}}} (q)$ and $\underline {{\psi _{{{\cal X}} \cup {{\cal Z}},{l_b}}}} (q)$ can be written as
\begin{small}
\begin{equation}
  \begin{array}{l}
    \overline {{\psi _{{{\cal X}} \cup {{\cal Y}},{l_a}}}} (q) =\\
    \;\;\;\; \sum\limits_{i \in {{\cal Z}} \cup {{\cal W}}} {\left\{ {\overline {{e_i}} ({s_{l_a}}) - \left( {{\bf{1}} - {{\bf{u}}_{{l_a}}}_{[{g_{l_a}}:q - 1]}} \right){{\underline {{{\bf{p}}_i}} }_{[{g_{l_a}}:q - 1]}}} \right\}} \\
    \;\;\;\;\;\;\;\;+ \sum\limits_{i \in {{\cal X}} \cup {{\cal Y}}} {{{\bf{u}}_{{l_a}}}_{[{g_{l_a}}:q - 1]}{{\overline {{{\bf{p}}_i}} }_{[{g_{l_a}}:q - 1]}}} 
    ,\end{array}\nonumber
\end{equation}
\begin{equation}
  \begin{array}{l}
    \underline {{\psi _{{{\cal X}} \cup {{\cal Z}},{l_b}}}} (q) =\\
    \;\;\;\; \sum\limits_{i \in {{\cal Y}} \cup {{\cal W}}} {\left\{ {\underline {{e_i}} ({s_{l_b}}) - \left( {{\bf{1}} - {{\bf{u}}_{{l_b}}}_{[{g_{l_b}}:q - 1]}} \right){{\overline {{{\bf{p}}_i}} }_{[{g_{l_b}}:q - 1]}}} \right\}} \\
    \;\;\;\;\;\;\;\;+ \sum\limits_{i \in {{\cal X}} \cup {{\cal Z}}} {{{\bf{u}}_{{l_b}}}_{[{g_{l_b}}:q - 1]}{{\underline {{{\bf{p}}_i}} }_{[{g_{l_b}}:q - 1]}}} 
    .\end{array}\nonumber
\end{equation}
\end{small}So the right-hand side of inequality \eqref{stt3res1} is
\begin{small}
\begin{equation}
  \label{final1}
  \begin{array}{l}
    \overline {{\varphi _{{{\cal X}} \cup {{\cal Y}},{l_a}}}} (q) - \underline {{\varphi _{{{\cal X}} \cup {{\cal Z}},{l_b}}}} (q) = \\
    \sum\limits_{i \in {{\cal X}}} {\left( {{{\bf{u}}_{{l_a}}}_{[{g_{l_a}}:q - 1]}{{\overline {{{\bf{p}}_i}} }_{[{g_{l_a}}:q - 1]}} - {{\bf{u}}_{{l_b}}}_{[{g_{l_b}}:q - 1]}{{\underline {{{\bf{p}}_i}} }_{[{g_{l_b}}:q - 1]}}} \right)} \\

     + \sum\limits_{i \in {{\cal Y}}} {\left\{ {{{\bf{u}}_{{l_a}}}_{[{g_{l_a}}:q - 1]}{{\overline {{{\bf{p}}_i}} }_{[{g_{l_a}}:q - 1]}} - \underline {{e_i}} (s_{l_b})} \right.} \\
     \;\;\;\;\;\;\;\;\left. { + \left( {{\bf{1}} - {{\bf{u}}_{{l_b}}}_{[{g_{l_b}}:q - 1]}} \right){{\overline {{{\bf{p}}_i}} }_{[{g_{l_b}}:q - 1]}}} \right\}\\

     + \sum\limits_{i \in {{\cal Z}}} {\left\{ {\overline {{e_i}} (s_{l_a}) - \left( {{\bf{1}} - {{\bf{u}}_{{l_a}}}_{[{g_{l_a}}:q - 1]}} \right){{\underline {{{\bf{p}}_i}} }_{[{g_{l_a}}:q - 1]}}} \right.} \\
     \;\;\;\;\;\;\;\;\left. { - {{\bf{u}}_{{l_b}}}_{[{g_{l_b}}:q - 1]}{{\underline {{{\bf{p}}_i}} }_{[{g_{l_b}}:q - 1]}}} \right\}\\

     + \sum\limits_{i \in {{\cal W}}} {\left\{ {\overline {{e_i}} (s_{l_a}) - \underline {{e_i}} (s_{l_b}) + \left( {{\bf{1}} - {{\bf{u}}_{{l_b}}}_{[{g_{l_b}}:q - 1]}} \right){{\overline {{{\bf{p}}_i}} }_{[{g_{l_b}}:q - 1]}}} \right.} \\
     \;\;\;\;\;\;\;\;\left. { - \left( {{\bf{1}} - {{\bf{u}}_{{l_a}}}_{[{g_{l_a}}:q - 1]}} \right){{\underline {{{\bf{p}}_i}} }_{[{g_{l_a}}:q - 1]}}} \right\}.

    \end{array}
\end{equation}
\end{small}

To complete the proof, the following definitions should be made.
\begin{definition}
  Say ${{\mathbf{u}}_{{l_a}}}$ and ${{\mathbf{u}}_{{l_b}}}$ overlap at $\theta$, if ${u_{{l_a}}}(\theta ) = {u_{{l_b}}}(\theta ) = 1$.
\end{definition}
\begin{definition}
  Say ${{\mathbf{u}}_{{l_a}}}$ and ${{\mathbf{u}}_{{l_b}}}$ do not overlap, if ${{\mathbf{u}}_{{l_a}}} + {{\mathbf{u}}_{{l_b}}} \le {\mathbf{1}}$.
\end{definition}

According to the above definitions, a new vector can be construncted in the following way: its $\theta$-th component is $1$, if and only if ${{\mathbf{u}}_{{l_a}}}$ and ${{\mathbf{u}}_{{l_b}}}$ overlap at $\theta$, otherwise the component is $0$. This vector, denoted by ${{\mathbf{u}}_{{l_0}}}$, corresponds to a new path $l_0$. Symmetrically, another vector can be constructed in the following way: its $\theta$-th component is $1$, if and only if ${u_{{l_a}}}(\theta ) =1 $ or $ {u_{{l_b}}}(\theta ) = 1$, otherwise the component is $0$. This vector, denoted as ${{\mathbf{u}}_{{l_\Sigma}}}$, also corresponds to a new path $l_\Sigma$. Another two vectors ${{\mathbf{u}}_{{l'_a}}}$ and ${{\mathbf{u}}_{{l'_b}}}$ (with their corresponding paths $l'_a$ and $l'b$) are defined by \eqref{deful1aul1b}. 
\begin{small}
\begin{equation}
  \label{deful1aul1b}
  \left\{ \begin{array}{c}
    {{\mathbf{u}}_{{l_a}}} := {{\mathbf{u}}_{{{l'}_a}}} + {{\mathbf{u}}_{{l_0}}},\\
    {{\mathbf{u}}_{{l_b}}}: = {{\mathbf{u}}_{{{l'}_b}}} + {{\mathbf{u}}_{{l_0}}}.
    \end{array} \right.
\end{equation}
\end{small}It is obvious that the above vectors satisfy
\begin{small}
\begin{equation}
  \label{overlap1}
  \left\{ \begin{array}{c}
    {{\mathbf{u}}_{{{l'}_a}}} + {{\mathbf{u}}_{{{l'}_b}}} \le {\mathbf{1}}({{\mathbf{u}}_{{{l'}_a}}} \text{ and } {{\mathbf{u}}_{{{l'}_b}}} \text{ do not overlap}),\\
    {{\mathbf{u}}_{{l_\Sigma }}} = {{\mathbf{u}}_{{{l'}_a}}} + {{\mathbf{u}}_{{l_b}}} = {{\mathbf{u}}_{{{l'}_b}}} + {{\mathbf{u}}_{{l_a}}}.
    \end{array} \right.
\end{equation}
\end{small}

Assume that $g_{l_a} \le g_{l_b}$ without loss of generality, so ${{\mathbf{u}}_{{l_a}}}$ and ${{\mathbf{u}}_{{l_b}}}$ must not overlap in $[ g_{l_a}, g_{l_b}-1] $. Hence, the equalities in \eqref{overlap2} hold.
\begin{small}
\begin{equation}
  \label{overlap2}
  \left\{ \begin{array}{l}
    {g_{{{l'}_a}}} = {g_{{l_a}}},{{\bf{u}}_{{l_0}}}_{[{g_{{l_a}}}:{g_{{l_b}}} - 1]} = {\bf{0}},\\
    {{\bf{u}}_{{l_0}}}_{[{g_{{l_a}}}:q - 1]} = {{\bf{u}}_{{l_0}}}_{[{g_{{l_b}}}:q - 1]},\\
     = {{\bf{u}}_{{l_a}}}_{[{g_{{l_a}}}:q - 1]} - {{\bf{u}}_{{{l'}_a}}}_{[{g_{{l_a}}}:q - 1]},\\
     = {{\bf{u}}_{{l_b}}}_{[{g_{{l_b}}}:q - 1]} - {{\bf{u}}_{{{l'}_b}}}_{[{g_{{l_b}}}:q - 1]}
    .\end{array} \right.
\end{equation}
\end{small}Because the number of non-zero nodes on $l_a$ and $l_b$ are both odd, then ${u_{{l_a}}}(q)$ and ${u_{{l_b}}}(q)$ must be $1$. Therefore, $l_a$ and $l_b$ must overlap at $q$. The number of non-zero nodes on $l'_a$ and $l'_b$ are both even because ${u_{{l'_a}}}(q)={u_{{l'_b}}}(q)=0$. Hence, the left-hand side of inequality \eqref{stt3res1} is equivalent to
\begin{small}
\begin{equation}
  \begin{array}{l}
    {\mathbf{u}}_{{{l'}_a}}{\mathbf{P}}(q) + \sum\limits_{i \in {{\mathcal Z}}} {{e_i}( t )} 
     - \left( {{\mathbf{u}}_{{{l'}_b}}{\mathbf{P}}(q) + \sum\limits_{i \in {{\mathcal Y}}} {{e_i}( t )} } \right)
    .\end{array}\nonumber
\end{equation}
\end{small}There exist
\begin{small}
\begin{equation}
  \label{alrhave1}
  {\mathbf{u}}_{{{l'}_a}}{\mathbf{P}}(q) + \sum\limits_{i \in {{\mathcal Z}}} {{e_i}( t )}  \le \overline {{\psi _{{{\mathcal Z}},{{l'}_a}}}} (q),
\end{equation}
\begin{equation}
  \label{alrhave2}
  {\mathbf{u}}_{{{l'}_b}}{\mathbf{P}}(q) + \sum\limits_{i \in {{\mathcal Y}}} {{e_i}( t )}  \ge \underline {{\psi _{{{\mathcal Y}},{{l'}_b}}}} (q)
\end{equation}
\end{small}where the left-hand sides are
\begin{small}
\begin{equation}
  \begin{array}{c}
    \overline {{\varphi _{{{\cal Z}},{{l'}_a}}}} (q) = \sum\limits_{i \in {{\cal Z}}} {\left\{ {\overline {{e_i}} ({s_{{{l'}_a}}}) - \left( {{\bf{1}} - {{\bf{u}}_{{{l'}_a}}}_{[{g_{{{l'}_a}}}:q - 1]}} \right){{\underline {{{\bf{p}}_i}} }_{[{g_{{{l'}_a}}}:q - 1]}}} \right\}} \\
     + \sum\limits_{i \in {{\cal Y}} \cup {{\cal W}} \cup {{\cal X}}} {{{\bf{u}}_{{{l'}_a}}}_{[{g_{{{l'}_a}}}:q - 1]}{{\overline {{{\bf{p}}_i}} }_{[{g_{{{l'}_a}}}:q - 1]}}} ,
    \end{array}\nonumber
\end{equation}
\begin{equation}
  \begin{array}{c}
    \underline {{\varphi _{{{\cal Y}},{{l'}_b}}}} (q) = \sum\limits_{i \in {{\cal Y}}} {\left\{ {\underline {{e_i}} ({s_{{{l'}_b}}}) - \left( {{\bf{1}} - {{\bf{u}}_{{{l'}_b}}}_{[{g_{{{l'}_b}}}:q - 1]}} \right){{\overline {{{\bf{p}}_i}} }_{[{g_{{{l'}_b}}}:q - 1]}}} \right\}} \\
     + \sum\limits_{i \in {{\cal Z}} \cup {{\cal W}} \cup {{\cal X}}} {{{\bf{u}}_{{{l'}_b}}}_{[{g_{{{l'}_b}}}:q - 1]}{{\underline {{{\bf{p}}_i}} }_{[{g_{{{l'}_b}}}:q - 1]}}} .
    \end{array}\nonumber
\end{equation}
\end{small}Therefore,
\begin{small}
\begin{equation}
  \label{final2}
  \begin{array}{l}
    \overline {{\varphi _{{{\cal Y}},{{l'}_a}}}} (q) - \underline {{\varphi _{{{\cal Z}},{{l'}_b}}}} (q) = \\
    \sum\limits_{i \in {{\cal X}}} {\left( {{{\bf{u}}_{{{l'}_a}}}_{[{g_{{{l'}_a}}}:q - 1]}{{\overline {{{\bf{p}}_i}} }_{[{g_{{{l'}_a}}}:q - 1]}} - {{\bf{u}}_{{{l'}_b}}}_{[{g_{{{l'}_b}}}:q - 1]}{{\underline {{{\bf{p}}_i}} }_{[{g_{{{l'}_b}}}:q - 1]}}} \right)} \\
     + \sum\limits_{i \in {{\cal Z}}} {\left\{ {\overline {{e_i}} ({s_{{{l'}_a}}}) - \left( {{\bf{1}} - {{\bf{u}}_{{{l'}_a}}}_{[{g_{{{l'}_a}}}:q - 1]}} \right){{\underline {{{\bf{p}}_i}} }_{[{g_{{{l'}_a}}}:q - 1]}}} \right.} \\
     \;\;\;\;\;\;\;\;\left. { - {{\bf{u}}_{{{l'}_b}}}_{[{g_{{{l'}_b}}}:q - 1]}{{\underline {{{\bf{p}}_i}} }_{[{g_{{{l'}_b}}}:q - 1]}}} \right\}\\
     + \sum\limits_{i \in {{\cal Y}}} {\left\{ {{{\bf{u}}_{{{l'}_a}}}_{[{g_{{{l'}_a}}}:q - 1]}{{\overline {{{\bf{p}}_i}} }_{[{g_{{{l'}_a}}}:q - 1]}} - \underline {{e_i}} ({s_{{{l'}_b}}})} \right.} \\
    \;\;\;\;\;\;\;\; + \left. {\left( {{\bf{1}} - {{\bf{u}}_{{{l'}_b}}}_{[{g_{{{l'}_b}}}:q - 1]}} \right){{\overline {{{\bf{p}}_i}} }_{[{g_{{{l'}_b}}}:q - 1]}}} \right\}\\

     + \sum\limits_{i \in {{\cal W}}} {\left( {{{\bf{u}}_{{{l'}_a}}}_{[{g_{{{l'}_a}}}:q - 1]}{{\overline {{{\bf{p}}_i}} }_{[{g_{{{l'}_a}}}:q - 1]}} - {{\bf{u}}_{{{l'}_b}}}_{[{g_{{{l'}_b}}}:q - 1]}{{\underline {{{\bf{p}}_i}} }_{[{g_{{{l'}_b}}}:q - 1]}}} \right)} .
    \end{array}
\end{equation}
\end{small}To prove that the inequality obtained after elimination \eqref{stt3res1} is implied by \eqref{alrhave1} and \eqref{alrhave2}, just prove
\begin{equation}
  \label{finalprove}
  \overline {{\psi _{{{\mathcal X}} \cup {{\mathcal Y}},{l_a}}}} (q) - \underline {{\psi _{{{\mathcal X}} \cup {{\mathcal Z}},{l_b}}}} (q) \ge \overline {{\psi _{{{\mathcal Y}},{{l'}_a}}}} (q) - \underline {{\psi _{{{\mathcal Z}},{{l'}_b}}}} (q).
\end{equation}
And this inequality is proved by comparing \eqref{final1} and \eqref{final2} item by item. Using \eqref{deful1aul1b}-\eqref{overlap2}, for the summation of $\mathcal X$, we have
\begin{equation}
  \begin{array}{l}
    \;\;\;\;{{\bf{u}}_{{l_a}}}_{[{g_{{l_a}}}:q - 1]}{\overline {{{\bf{p}}_i}} _{[{g_{{l_a}}}:q - 1]}} - {{\bf{u}}_{{l_b}}}_{[{g_{{l_b}}}:q - 1]}{\underline {{{\bf{p}}_i}} _{[{g_{{l_b}}}:q - 1]}}\\
    - \left( {{{\bf{u}}_{{{l'}_a}}}_{[{g_{{{l'}_a}}}:q - 1]}{{\overline {{{\bf{p}}_i}} }_{[{g_{{{l'}_a}}}:q - 1]}} - {{\bf{u}}_{{{l'}_b}}}_{[{g_{{{l'}_b}}}:q - 1]}{{\underline {{{\bf{p}}_i}} }_{[{g_{{{l'}_b}}}:q - 1]}}} \right)\\
     = {{\bf{u}}_{{l_0}}}_{[{g_{{l_a}}}:q - 1]}{\overline {{{\bf{p}}_i}} _{[{g_{{l_a}}}:q - 1]}} - {{\bf{u}}_{{l_0}}}_{[{g_{{l_b}}}:q - 1]}{\underline {{{\bf{p}}_i}} _{[{g_{{l_b}}}:q - 1]}} \\
     = {{\bf{u}}_{{l_0}}}_{[{g_{{l_b}}}:q - 1]}\left( {{{\overline {{{\bf{p}}_i}} }_{[{g_{{l_b}}}:q - 1]}} - {{\underline {{{\bf{p}}_i}} }_{[{g_{{l_b}}}:q - 1]}}} \right)\\
     \ge 0 \text{  (Hypo 1)}
    ;\end{array}\nonumber
\end{equation}for the summation of $\mathcal Y$, we have
\begin{small}
\begin{equation}
  \begin{array}{l}
    \begin{array}{l}
      {{\bf{u}}_{{l_a}}}_{[{g_{{l_a}}}:q - 1]}{\overline {{{\bf{p}}_i}} _{[{g_{{l_a}}}:q - 1]}} - \underline {{e_i}} ({s_{{l_b}}})\\
      \;\;\;\;+ \left( {{\bf{1}} - {{\bf{u}}_{{l_b}}}_{[{g_{{l_b}}}:q - 1]}} \right){\overline {{{\bf{p}}_i}} _{[{g_{{l_b}}}:q - 1]}}\\
      \;\;\;\;\;\;\;\;- {{\bf{u}}_{{{l'}_a}}}_{[{g_{{{l'}_a}}}:q - 1]}{\overline {{{\bf{p}}_i}} _{[{g_{{{l'}_a}}}:q - 1]}} + \underline {{e_i}} ({s_{{{l'}_b}}})\\
      \;\;\;\;\;\;\;\;\;\;\;\;- \left( {{\bf{1}} - {{\bf{u}}_{{{l'}_b}}}_{[{g_{{{l'}_b}}}:q - 1]}} \right){\overline {{{\bf{p}}_i}} _{[{g_{{{l'}_b}}}:q - 1]}}
      \end{array}\\
     = {{\bf{u}}_{{l_0}}}_{[{g_{{l_a}}}:q - 1]}{\overline {{{\bf{p}}_i}} _{[{g_{{l_a}}}:q - 1]}} - {{\bf{u}}_{{l_0}}}_{[{g_{{l_b}}}:q - 1]}{\overline {{{\bf{p}}_i}} _{[{g_{{l_b}}}:q - 1]}}\\
     \;\;\;\;- \left( {\underline {{e_i}} ({s_{{l_b}}}) - \underline {{e_i}} ({s_{{{l'}_b}}}) - {\bf{1}}{{\overline {{{\bf{p}}_i}} }_{[{g_{{l_b}}}:{g_{{{l'}_b}}} - 1]}}} \right)\\
     = \underline {{e_i}} ({s_{{{l'}_b}}}) - \underline {{e_i}} ({s_{{l_b}}}) + \sum\limits_{t = {s_{{{l'}_b}}} + 1}^{{s_{{l_b}}}} {\overline {{p_i}} (t)} \\
     \ge 0 \text{  (Hypo 2)}
    ;\end{array}\nonumber
\end{equation}
\end{small}for $\mathcal Z$, we have
\begin{small}
\begin{equation}
  \begin{array}{l}
    \overline {{e_i}} ({s_{{l_a}}}) - \left( {{\bf{1}} - {{\bf{u}}_{{l_a}}}_{[{g_{{l_a}}}:q - 1]}} \right){\underline {{{\bf{p}}_i}} _{[{g_{{l_a}}}:q - 1]}}\\
    \;\;\;\;- {{\bf{u}}_{{l_b}}}_{[{g_{{l_b}}}:q - 1]}{\underline {{{\bf{p}}_i}} _{[{g_{{l_b}}}:q - 1]}}\\
    \;\;\;\;\;\;\;\;-\overline {{e_i}} ({s_{{{l'}_a}}}) + \left( {{\bf{1}} - {{\bf{u}}_{{{l'}_a}}}_{[{g_{{{l'}_a}}}:q - 1]}} \right){\underline {{{\bf{p}}_i}} _{[{g_{{{l'}_a}}}:q - 1]}}\\
    \;\;\;\;+ {{\bf{u}}_{{{l'}_b}}}_{[{g_{{{l'}_b}}}:q - 1]}{\underline {{{\bf{p}}_i}} _{[{g_{{{l'}_b}}}:q - 1]}}
    \\
     = {{\bf{u}}_{{l_0}}}_{[{g_{{{l'}_a}}}:q - 1]}{\underline {{{\bf{p}}_i}} _{[{g_{{{l'}_a}}}:q - 1]}} - {{\bf{u}}_{{l_0}}}_{[{g_{{l_b}}}:q - 1]}{\underline {{{\bf{p}}_i}} _{[{g_{{{l'}_b}}}:q - 1]}}\\
     = 0
    ;\end{array}\nonumber
\end{equation}\end{small}and for $\mathcal W$, we have
\begin{small}
\begin{equation}
  \begin{array}{l}
    \overline {{e_i}} ({s_{{l_a}}}) - \underline {{e_i}} ({s_{{l_b}}})\\
    \;\;\;\;+ \left( {{\bf{1}} - {{\bf{u}}_{{l_b}}}_{[{g_{{l_b}}}:q - 1]}} \right){\overline {{{\bf{p}}_i}} _{[{g_{{l_b}}}:q - 1]}} - \left( {{\bf{1}} - {{\bf{u}}_{{l_a}}}_{[{g_{{l_a}}}:q - 1]}} \right){\underline {{{\bf{p}}_i}} _{[{g_{{l_a}}}:q - 1]}}\\
    \;\;\;\;\;\;\;\;- \left( {{{\bf{u}}_{{{l'}_a}}}_{[{g_{{{l'}_a}}}:q - 1]}{{\overline {{{\bf{p}}_i}} }_{[{g_{{{l'}_a}}}:q - 1]}} - {{\bf{u}}_{{{l'}_b}}}_{[{g_{{{l'}_b}}}:q - 1]}{{\underline {{{\bf{p}}_i}} }_{[{g_{{{l'}_b}}}:q - 1]}}} \right)\\
     = \overline {{e_i}} ({s_{{l_a}}}) - \underline {{e_i}} ({s_{{l_b}}}) + {\left( {{\bf{1}} - {{\bf{u}}_{{l_b}}} - {{\bf{u}}_{{{l'}_a}}}} \right)_{[{g_{{l_b}}}:q - 1]}}{\left( {\overline {{{\bf{p}}_i}}  - \underline {{{\bf{p}}_i}} } \right)_{[{g_{{l_b}}}:q - 1]}}\\
     \;\;\;\;- {{\bf{u}}_{{{l'}_a}}}_{[{g_{{l_a}}}:{g_{{l_b}}} - 1]}{\overline {{{\bf{p}}_i}} _{[{g_{{l_a}}}:{g_{{l_b}}} - 1]}} - \left( {{\bf{1}} - {{\bf{u}}_{{l_a}}}_{[{g_{{l_a}}}:{g_{{l_b}}} - 1]}} \right){\underline {{{\bf{p}}_i}} _{[{g_{{l_a}}}:{g_{{l_b}}} - 1]}}\\
     = \overline {{e_i}} ({s_{{l_a}}}) - \underline {{e_i}} ({s_{{l_b}}}) + {\bf{1}}{\underline {{{\bf{p}}_i}} _{[{g_{{l_a}}}:{g_{{l_b}}} - 1]}}\\
     \;\;\;\;+ \left( {{\bf{1}} - {{\bf{u}}_{{l_\Sigma }}}_{[{g_{{l_a}}}:q - 1]}} \right){\left( {\overline {{{\bf{p}}_i}}  - \underline {{{\bf{p}}_i}} } \right)_{[{g_{{l_a}}}:q - 1]}}\\
     \ge 0{\text{  }}({\text{Hypo 1 and Hypo 2}})
    .\end{array}\nonumber
\end{equation} \end{small}Therefore, the inequality \eqref{finalprove} holds, and this completes the proof of the situation where $I_{l_a}=I_{l_b}=1$. 

For the situation where $I_{l_a}=1$ and $I_{l_b}=0$, the elimination process will produce
\begin{small}
\begin{equation}
  \begin{array}{l}
    ({\mathbf{u}}_{{l_a}} + {\mathbf{u}}_{{l_b}}){\mathbf{P}}(q)+\left( { - \sum\limits_{i \in {{\mathcal Y}}} {{e_i}( t )}  + \sum\limits_{i \in {{\mathcal Z}}} {{e_i}( t )} } \right)\\
     \le \overline {{\gamma _{{{\mathcal X}} \cup {{\mathcal Y}},{l_a}}}} (q) + \overline {{\gamma _{{{\mathcal X}} \cup {{\mathcal Z}},{l_b}}}} (q)
    .\end{array}
\end{equation} 
\end{small}According to the definition of $l_\Sigma$ and $l_0$, the above inequality is equivalent to
\begin{small}
\begin{equation}
  \label{stt2}
  \begin{array}{l}
    ({\mathbf{u}}_{{l_\Sigma }} + {\mathbf{u}}_{{l_0}}){\mathbf{P}}(q)+ \left( { - \sum\limits_{i \in {{\mathcal Y}}} {{e_i}( t )}  + \sum\limits_{i \in {{\mathcal Z}}} {{e_i}( t )} } \right)\\
     \le \overline {{\gamma _{{{\mathcal X}} \cup {{\mathcal Y}},{l_a}}}} (q) + \overline {{\gamma _{{{\mathcal X}} \cup {{\mathcal Z}},{l_b}}}} (q)
    \end{array}
\end{equation} 
\end{small}
where the number of non-zero nodes on $l_\Sigma$ is odd, and the number of non-zero nodes on $l_0$ is even. Consider adding the following inequalities, the redundancy of inequality \eqref{stt2} can be proved.
\begin{small}
\begin{equation}
  \left\{ \begin{array}{l}
    {\mathbf{u}}_{{l_\Sigma }}{\mathbf{P}}(q) - \sum\limits_{i \in {{\mathcal Y}}} {{e_i}( t ) \le \overline {{\psi _{{{\mathcal Y}},{l_\Sigma }}}} (q)} \\
    {\mathbf{u}}_{{l_0}}{\mathbf{P}}(q) + \sum\limits_{i \in {{\mathcal Z}}} {{e_i}( t ) \le \overline {{\psi _{{{\mathcal Z}},{l_0}}}} (q)} 
    \end{array} \right.
\end{equation} 
\end{small}
For the situation where $I_{l_a}=0$ and $I_{l_b}=1$, the redundancy can be proved by adding the following inequalities.
\begin{small}
\begin{equation}
  \left\{ \begin{array}{l}
    {\mathbf{u}}_{{l_\Sigma }}{\mathbf{P}}(q) + \sum\limits_{i \in {{\mathcal Z}}} {{e_i}( t )}  \le \overline {{\psi _{{{\mathcal Z}},{l_\Sigma }}}} (q)\\
    {\mathbf{u}}_{{l_0}}{\mathbf{P}}(q) + \sum\limits_{i \in {{\mathcal Y}}} {{e_i}( t )}  \le \overline {{\psi _{{{\mathcal Y}},{l_0}}}} (q)
    \end{array} \right.
\end{equation} 
\end{small}
For the situation where $I_{l_a}=I_{l_b}=0$, the number of non-zero elements of $l'_a$ and $l'_b$ are also both even. Therefore, the redundancy can be proved by subtracting the following inequalities.
\begin{small}
\begin{equation}
  \left\{ \begin{array}{l}
    {\mathbf{u}}_{{{l'}_a}}{\mathbf{P}}(q) + \sum\limits_{i \in {{\mathcal Y}}} {{e_i}( t )}  \le \overline {{\psi _{{{\mathcal Y}},{{l'}_a}}}} (q)\\
    {\mathbf{u}}_{{{l'}_b}}{\mathbf{P}}(q) + \sum\limits_{i \in {{\mathcal Z}}} {{e_i}( t )}  \ge \underline {{\psi _{{{\mathcal Z}},{{l'}_b}}}} (q)
    \end{array} \right.
\end{equation} 
\end{small}
This completes the proof of Theorem \ref{stt3redundancy}.


\ifCLASSOPTIONcaptionsoff
  \newpage
\fi
\small
\bibliographystyle{IEEEtran}

\bibliography{ref}





\end{document}